\documentclass[aip,jcp,superscripedaddress]{revtex4-1}
\usepackage{amsmath}
\usepackage{latexsym}
\usepackage{float}
\usepackage{amssymb}
\usepackage{graphicx}
\usepackage{epsfig}
\usepackage{color}

\definecolor{red}{rgb}{1.0,0.0,0.0}

\begin{document}

\title{Computer simulation of bottle brush polymers with flexible
backbone: Good solvent versus Theta solvent conditions}

\author{Panagiotis E Theodorakis}
\email{panagiotis.theodorakis@univie.ac.at}
\affiliation{Institut f\"ur Physik, Johannes Gutenberg-Universit\"at Mainz,\\
Staudinger Weg 7, D-55099 Mainz, Germany}
\affiliation{Faculty of Physics, University of Vienna,
Boltzmanngasse 5, A-1090 Vienna, Austria}
\affiliation{Institute for Theoretical Physics and Center for
Computational Materials Science (CMS), Technical University of
Vienna, Hauptstra{$\beta$}e 8-10, A-1040 Vienna, Austria}
\affiliation{Vienna Computational Materials' Laboratory,  
Sensengasse 8/12, A-1090 Vienna, Austria}
\author{Hsiao-Ping Hsu}
\email{hsu@uni-mainz.de}
\affiliation{Institut f\"ur Physik, Johannes Gutenberg-Universit\"at Mainz,\\
 Staudinger Weg 7, D-55099 Mainz, Germany}
\author{Wolfgang Paul}
\email{wolfgang.paul@physik.uni-halle.de}
\affiliation{Theoretische Physik, Martin Luther Universit\"at \\
Halle-Wittenberg, von Seckendorffplatz 1, 06120 Halle, Germany}
\author{Kurt Binder}
\email{kurt.binder@uni-mainz.de}
\affiliation{Institut f\"ur Physik, Johannes Gutenberg-Universit\"at Mainz,\\
 Staudinger Weg 7, D-55099 Mainz, Germany}

\date{\today}
\begin{abstract}
By Molecular Dynamics simulation of a coarse-grained bead-spring type
model for a cylindrical molecular brush with a backbone chain of $N_b$ effective monomers
to which with grafting density $\sigma$ side chains 
with $N$ effective monomers are tethered, several characteristic
length scales are studied for variable solvent quality. Side chain
lengths are in the range $5 \le N \le 40$, backbone chain lengths are in the range
$50 \le N_b \le 200$, and we perform a comparison to results for the bond
fluctuation model on the simple cubic lattice (for which much longer
chains are accessible, $N_b \le 1027$, and which
corresponds to an athermal, very good, solvent). We obtain linear
dimensions of side chains and the backbone chain and discuss their
$N$-dependence in terms of power laws and the associated effective
exponents. We show that even at the Theta point the side chains are 
considerably stretched, their linear dimension depending on the solvent
quality only weakly. Effective persistence lengths are extracted both
from the orientational correlations and from the backbone end-to-end
distance; 
it is shown that different measures of the persistence length (which
would all agree for Gaussian chains) are not mutually consistent with
each other, and depend distinctly both on $N_b$ and the solvent
quality. A brief discussion of pertinent experiments is given.
\end{abstract}

\maketitle

\section{Introduction}
Macromolecules which consist of a ``backbone'' polymer,
to which flexible or stiff side chains are grafted, the so-called
``bottle brush polymers'', find very great interest recently (see
Refs~\cite{1,2,3,4,5,6} for reviews). Varying the chemical
nature of both backbone chain and side chains, their chain
lengths $(N_b, N)$ and the grafting density $\sigma$, the structure
of these cylindrical molecular brushes can be widely varied.
I.e., their local ``thickness" as measured by the cross-sectional
radius $R_c$ or linear dimensions of individual side chains can
be varied as well as their local ``stiffness'', traditionally
measured by ``the'' persistence length $l_p$~\cite{6,7,8,9},
and their effective contour length $L_c$.
We here use quotation marks with respect to ``the'' persistence 
length, because there is evidence, at least for the case of
very good solvent conditions, that a unique persistence length
measuring the ``intrinsic'' stiffness of a polymer cannot be defined
in the standard fashion~\cite{10,11}. Now an intriguing observation~\cite{12}
is the sensitivity of the large-scale structure of these bottle brush
polymers to solvent quality: one finds a thermally induced collapse of
single macromolecules from cylindrical brushes to spheres, in a very small
temperature range, and it is speculated that these bottle brushes could be 
useful as building blocks of ``soft nanomachines''~\cite{12}. We also
note that biopolymers with comb-like architecture are ubiquitous  
in nature (such as proteoglycans~\cite{13}, or the aggrecane molecules that
play a role in the soft lubricating layers in human joints~\cite{14}, etc.),
and probably in this context temperature and/or solvent quality (or pH
value) are relevant parameters as well.

 In view of these facts, a comprehensive clarification of how the 
properties of bottle brush polymers depend on solvent quality clearly
would be interesting. Although there are occasional experimental reports,
how particular linear dimensions of these polymers scale in various
solvents (e.g.~\cite{15,16,17,18}), we are not aware of a systematic study
of this problem. While work based on self-consistent field theory (SCFT)
predicted already very early on~\cite{19} that 
the side chain gyration radius $R_{gs}$ scales as $R_{gs} \propto N^{3/4}$ for 
good solvents and $R_{gs} \propto N^{2/3}$ for Theta solvents, there is now
evidence from experiment, simulation and theory that these power laws
apply if at all only for side chain length $N$ of the order of $10^3$,
which are of no practical relevance: Experiments have only studied the
range $N<10^2$, and the range accessible in 
simulations~\cite{3,6,10,20,20a,21,22,22a,23,24} is similarly restricted.
Numerical modeling applying the Scheutjens-Fleer version of 
SCFT has given
clear evidence~\cite{25} that even within this mean-field approach $N \approx 10^3$ is needed to reach the 
regime where the predicted power laws~\cite{19,26}
apply. 
{While the investigation of the scaling (self-similar)
properties of bottle brushes with extremely long backbone chains and very
long side chains may be a challenging theoretical problem, it is of
little relevance for the experimentally accessible systems, and clearly
not in the focus of the present paper.}
Both SCFT theories~\cite{19,25} and scaling 
theories~\cite{27} consider ideal chains aiming to describe the behavior 
at the Theta temperature, while
previous simulations have almost exclusively considered the good solvent
case only. Notable exceptions are studies of globule formation of
bottle brushes with very short side chains ($N=4-12$) under poor solvent
conditions~\cite{28} and the study of microphase separation of bottle
brushes with straight backbones in poor solvent~\cite{29,30}.

 The present work intends to make a contribution to close this
gap, by presenting a simulation study of a coarse-grained bead-spring
type model of bottle brush polymers where the solvent quality is varied
from very good solvent conditions to the Theta point regime. In the next
section, we shall describe the model and simulation technique, while in
section III we present our numerical results. In Section IV, a summary is 
given, as well as an outlook on pertinent experiments.

\section{Model and Simulation Method}

Extending our previous work on the simulation of bottle brush polymers
with rigid backbones~\cite{29,30}, we describe both the backbone chain and
the side chains by a bead-spring model~\cite{31,32,33,34}, where all beads
interact with a truncated and shifted Lennard-Jones potential $U_{\rm LJ}(r)$
and nearest neighbors bonded together along a chain also experience the
finitely extensible nonlinear elastic potential $U_{\rm FENE}(r)$,
$r$ being the distance between the beads. Thus
\begin{equation}
  U_{\rm LJ}(r)=4 \varepsilon_{\rm LJ} 
\left[\left(\frac{\sigma_{\rm LJ}}{r}\right)^{12}-
\left(\frac{\sigma_{\rm LJ}}{r}\right)^6 \right]+C \, ,\qquad  r\le r_c \, ,
\label{eq1}
\end{equation}
while $U_{\rm LJ}(r>r_c)=0$, and where $r_c=2.5\sigma_{\rm LJ}$. 
The constant $C$ is defined such that $U_{\rm LJ}(r=r_c)$ is continuous at 
the cutoff. Henceforth units are chosen such that $\varepsilon_{\rm LJ}=1$,
$\sigma_{\rm LJ}=1$, the Boltzmann constant $k_B=1$, and in addition 
also the mass $m_{\rm LJ}$
of beads is chosen to be unity. The potential Eq.~(\ref{eq1}) acts between
any pair of beads, irrespective of whether they are bonded or not.
For bonded beads additionally the potential $U_{\rm FENE}(r)$ acts,
\begin{equation}
  U_{\rm FENE}(r)=-\frac{1}{2} kr_0^2 \ln \left[1-
\left(\frac{r}{r_0}\right)^2\right] \, , \qquad 
0 < r \le r_0 \, ,
\label{eq2}
\end{equation}
while $U_{\rm FENE}(r>r_0)=\infty$, and hence $r_0$ is the maximal
distance that bonded beads can take. We use the standard choice~\cite{34}
$r_0=1.5$ and $k=30$.
{
Related models have been used with great success to study the
glassification of polymer melts formed from short chains~\cite{x}
and to study the effects of solvent quality of polymer brushes on flat planar
substrates~\cite{x1}. For such models of brushes on a planar substrate, the
implicit solvent model (Eq.~(1)) has been compared with models using
explicit solvent molecules, and it was found that the results are
very similar.}

  Note that in our model there is no difference in interactions, irrespective
of whether the considered beads are effective monomers of the backbone or
of the side chains, implying that the polymer forming the backbone is 
either chemically identical to the polymers that are tethered as side chains 
to the backbone, or at least on coarse-grained length scales as considered
here the backbone and side chain polymers are no longer distinct.
There is also no difference between the bond linking the first 
monomer of a side chain to a monomer of the backbone and bonds between any
other pairs of bonded monomers. Of course, our study does not address any 
effects due to a particular chemistry relating to the synthesis of these
bottle brush polymers, but, as usually done~\cite{34,35,36} we address
universal features of the conformational properties of these macromolecules. 

  There is one important distinction relating to our previous work~\cite{29,30}
on bottle brush polymers with rigid backbones: following 
Grest and Murat~\cite{33}, there the backbone was taken as an infinitely thin 
straight line in continuous space, thus allowing arbitrary values of the
distances between neighboring grafting sites, and hence the grafting 
density $\sigma$ could be continuously varied. For the present model,
where we disregard any possible quenched disorder resulting from the 
grafting process, 
of course, the grafting density $\sigma$ is quantized: we denote here by
$\sigma=1$ the case that every backbone monomer carries a side chain,
$\sigma=0.5$ means that every second backbone monomer has a side chain,
etc. Chain lengths of side chains were chosen as $N=5$, $10$, $20$, and $40$,
while backbone chain lengths were chosen as $N_b=50$, $100$, and $200$, 
respectively. 

  It is obvious, of course, that for such short side chain lengths any
interpretation of characteristic lengths in terms of power laws, such as 
$R_c \propto N^{\nu_{\rm eff}}$, is a delicate matter, $\nu_{\rm eff}$
being an ``effective exponent'' and characterizes only the specified
range of rather small values of $N$, and not the limit $N \rightarrow \infty$
considered by most theories~\cite{2,19,25,26,27}. 
{
Thus, the actual value of $\nu_{\rm eff}$ is of limited interest,
it only gives an indication to which part of an extended crossover region
the data belong.}
However, we emphasize
that (i) our range of $N$ nicely corresponds to the range available in
experiments~\cite{1,4,15,16,17,18,24,37,38,39}. (ii) The analysis in terms
of power laws with effective exponents is a standard practice of 
experimentalists in this context (e.g.~\cite{1,17}).

  We recall that for linear chains the Theta temperature for the present
{(implicit solvent)} model has been roughly 
estimated~\cite{40} as $T_\theta \approx 3.0$
(note, however, that there is still some uncertainty about the
precise value of $T_\theta$: for a similar model~\cite{41} the correct
value of $T_\theta$, $T_\theta \approx 3.18$ in this case, could only be 
established for chain lengths exceeding $N=200$). Thus, in the present
work we have thoroughly studied the temperature range $3.0 \le T \le 4.0$.
From previous work~\cite{42} on rather long chains in polymer brushes
on flat surfaces, using the same model {(}
Eqs.~(\ref{eq1}), (\ref{eq2}){)}
to describe the interactions, it is known that for $T=4.0$
one finds a behavior characteristic for (moderately) good solvents.
Very good solvent conditions could be obtained from a slightly different
model that has extensively been studied for standard polymer 
brushes~\cite{34,43}, where in Eq.~(\ref{eq1}) the cutoff is chosen
to coincide with the minimum of the potential, $r_c=2^{1/6} \sigma_{\rm LJ}$
(and then also $T=1$ can be chosen for this essentially athermal model).
Rather than carrying out simulations for bottle brushes using this model,
we found it more appropriate to compare to 
{the athermal version of} the bond fluctuation
model on the simple cubic lattice, which 
describes very good solvent conditions~\cite{44,45} and has been used in
our earlier work~\cite{6,10,24}. The use of this model has several advantages:
(i) due to the fact that excluded volume constraints can be monitored via
the occupancy of lattice sites, a very efficient implementation of the
pivot algorithm in this Monte Carlo approach has become possible~\cite{46}.
Therefore, very large bottle brush polymers can be equilibrated, up to 
$N_b=1027$, a task that is very difficult to achieve by Molecular
Dynamics (MD) methods (ii) The extent to which the bond fluctuation model
and MD results agree (for comparable choices of parameters) yields some
insight to what extent the preasymptotic regime that we study is 
model dependent. Of course, a truly universal behavior (apart from 
amplitude prefactors) can only be expected for the asymptotic regime where 
the side chain length $N \rightarrow \infty$, that is not accessible in 
simulations or experiments. {One could ask why we do not use 
versions of the bond fluctuation model
where an effective attraction between monomers is included~\cite{xb}
to study the effect of variable solvent quality in the framework of this
model. The reason is that for temperatures slightly below the Theta
temperature already practically frozen configurations of monomers occur,
with several monomers next to each other blocking any possibility to move.
Thus the convergence towards equilibrium then is extremely slow.}

  However, since the application of {the athermal
version of} the bond fluctuation model 
(BFM) to the
simulation of bottle brush polymers is 
well documented 
{(Hsu and Paul~\cite{46} have given a careful discussion 
of the effort needed for the BFM to sample equilibrium properties.)}
in the recent literature~\cite{10,24,46}, we do not
give any details here. 

In the MD simulation, the positions $\vec{r}_i(t)$
of the effective monomers with label $i$ evolve in time $t$
according to Newton's equation of motion, amended by the Langevin
thermostat~\cite{31,32,33,34,35,36}
{\begin{equation}
  m_{\rm LJ} \frac{d^2 \vec{r}_i(t)}{dt^2} = -\nabla U_i \left(
\left\{ \vec{r}_j(t)\right\} \right)-\gamma \frac{d \vec{r}_i}{dt}
+\vec{\Gamma}_i(t) \, ,
\label{eq3}
\end{equation}}
where $U_i\left(\left\{ \vec{r}_j(t)\right\}\right)$ is the total potential
acting on the $i$'th bead due to its interactions with the other beads
at sites $\left\{ \vec{r}_j(t)\right\}$, $\gamma$ is the friction coefficient,
and $\vec{\Gamma}_i(t)$ the associated random force. The latter is related to
$\gamma$ by the fluctuation-dissipation relation
{\begin{equation}
 \langle \vec{\Gamma}_i(t) \cdot \vec{\Gamma}_j(t) \rangle
= 6 k_BT \frac{\gamma}{m_{\rm LJ}} \delta_{ij} \delta(t-t') \, .
\label{eq4}
\end{equation}}
Following previous work~\cite{29,30,31,32,33,34} we choose $\gamma=0.5$, the 
MD time unit
\begin{equation}
  \tau_{\rm LJ} =\left(\frac{m_{\rm LJ} \sigma^2_{\rm LJ}}
{\varepsilon_{\rm LJ}}\right)^{1/2}
\label{eq5}
\end{equation}
being also unity, for our choice of units. Eq.~(\ref{eq3}) was integrated
using the leap frog algorithm~\cite{47}, with a time step 
$\delta t=0.006 \tau_{\rm LJ}$, and utilizing the GROMACS package~\cite{48}.
For the calculation of properties of the bottle brushes typically $500$
statistically independent configurations are averaged over.

\begin{figure}
\begin{center}
(a)\includegraphics[scale=0.28,angle=270]{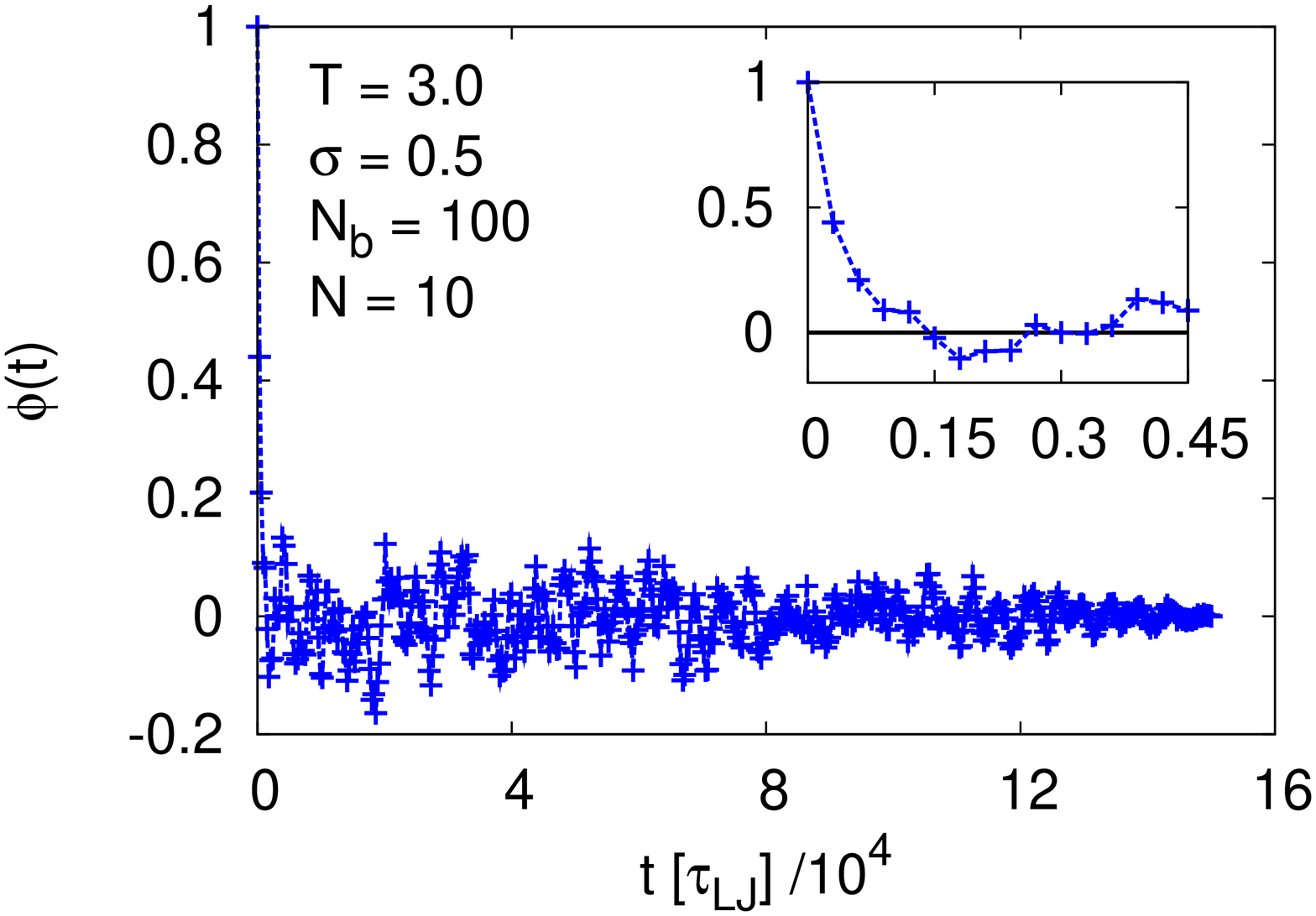} \hspace{0.8cm}
(b)\includegraphics[scale=0.28,angle=270]{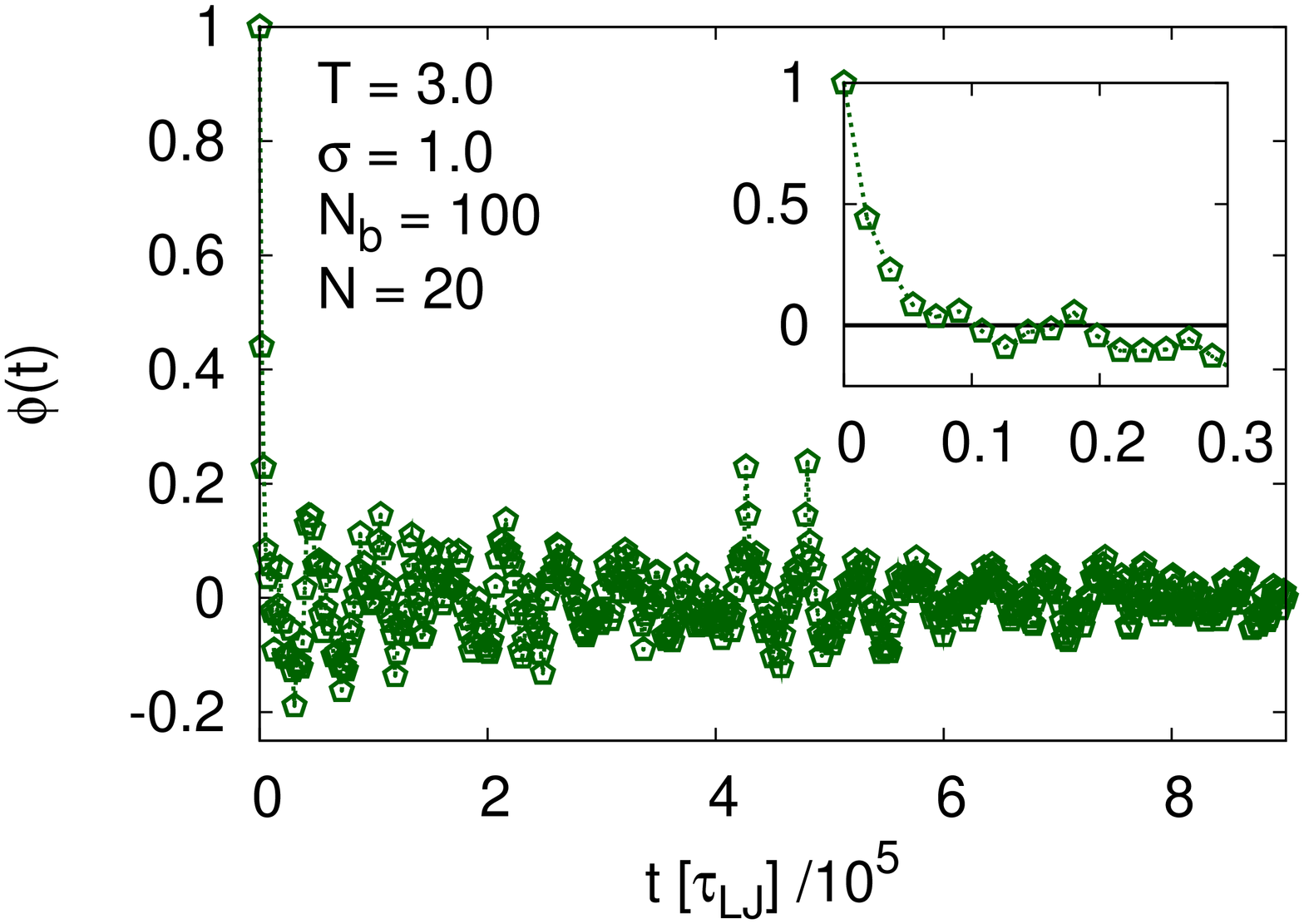}\\
(c)\includegraphics[scale=0.28,angle=270]{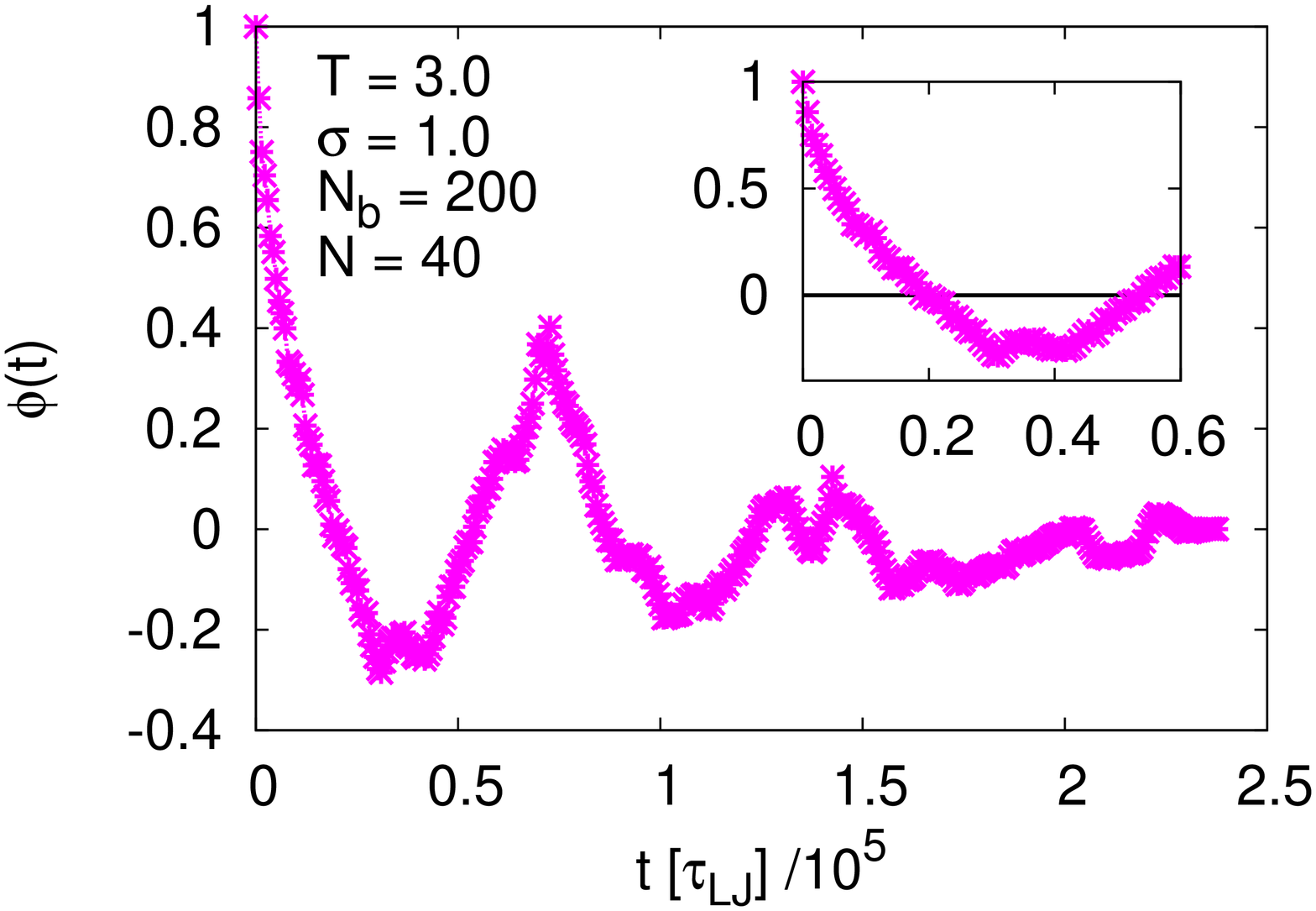} 
\caption{Plot of $\phi(t)$ (see Eq.~(\ref{eq6}))
versus the time $t$ in units of
$\tau_{\rm LJ}$.
Several cases are shown: $\sigma=0.5$, $N_b=100$, $N=10$, $T=3.0$ (a)
$\sigma=1.0$, $N_b=100$, $N=20$, $T=3.0$ (b); and $\sigma=1.0$, $N_b=200$, $N=40$,
$T=3.0$ (c). The fluctuations of $\phi(t)$ change sign initially at 
$t \approx 1500\tau_{\rm LJ}$ (a), $t \approx 10000 \tau_{\rm LJ}$, and 
$t \approx 20000\tau_{\rm LJ}$ (c), as shown in the inserts.}
\label{fig1}
\end{center}
\end{figure}

Of course, for bottle brushes with large $N_b$ equilibration
of the polymer conformations is a difficult problem. Since we expect
that end-to-end distance ${R}_e$ and gyration radius ${R}_g$
of the whole molecule belong to the slowest relaxing quantities,
we studied the autocorrelation function of $R_g^2$ to
test for equilibration,
\begin{equation}
  \phi(t)=\frac{\langle \left(R_g^2(t'+t)-\overline{R_g^2}\right)\left(
R_g^2(t')-\overline{R_g^2}\right)\rangle} {\langle \left(R^2_g(t')-
\overline{R_g^2}\right)^2\rangle}
\label{eq6}
\end{equation}
 Note that $\overline{R_g^2}$ means an average of $R_g^2(t')$ over the time
$t'$ in the MD trajectory (and the average $\langle \cdots \rangle$
is computed by averaging over $500$ statistically independent runs).
Despite a substantial investment of computer time, $\phi(t)$ still exhibits
significant fluctuations (remember that quantities such as ${R}_e$
and ${R}_g$ are known to exhibit a ``lack of self-averaging~\cite{49,50}'',
irrespective of how large $N_b$ is). Fig.~\ref{fig1} gives some examples for
$\phi(t)$. If the number of samples would be infinite and 
$\delta t \rightarrow 0$, we should expect a monotonous decay of $\phi(t)$
towards zero as $t$ becomes large. Due to the fact that the number of 
samples is not extremely large, and $\delta t$ not extremely small,
Fig.~\ref{fig1} gives clear evidence for noise that is still correlated.
We see that for small $N$ the noise amplitude starts out at $\pm 0.1$, and 
the time scale on which the fluctuations of $\phi(t)$ change sign is at
$t \approx 1500 \tau_{\rm LJ}$ in (a), $t \approx 10000 \tau_{\rm LJ}$ in (b), and 
$t \approx 20000 \tau_{\rm LJ}$ in (c). While in cases (a)(b) the (statistically
meaningful) initial decay of $\phi(t)$ occurs so fast that it 
can only be seen on a magnified abscissa scale (inserts),
we see that in (c) the initial decay is also much
slower, and the associated time scale is of the same order as the time over
which fluctuations are correlated. We have carefully considered $\phi(t)$
for all cases studied, and we have concluded that for our largest system
studied (shown in Fig.~\ref{fig1}(c), with a total number of $8\;000$
effective monomers) the statistical effort is not yet sufficient to allow
meaningful conclusions on the overall linear dimensions of the bottle
brush, while in all other cases the effort was judged to be sufficient.
{The damped oscillatory character of the relaxation seen 
particularly
in Fig.~\ref{fig1}(c) could be a matter of concern; we attribute
this relaxation behavior to particular slow collective motions
(breathing type modes) of the chain.}

\begin{figure}
\begin{center}
\includegraphics[scale=0.45,angle=0]{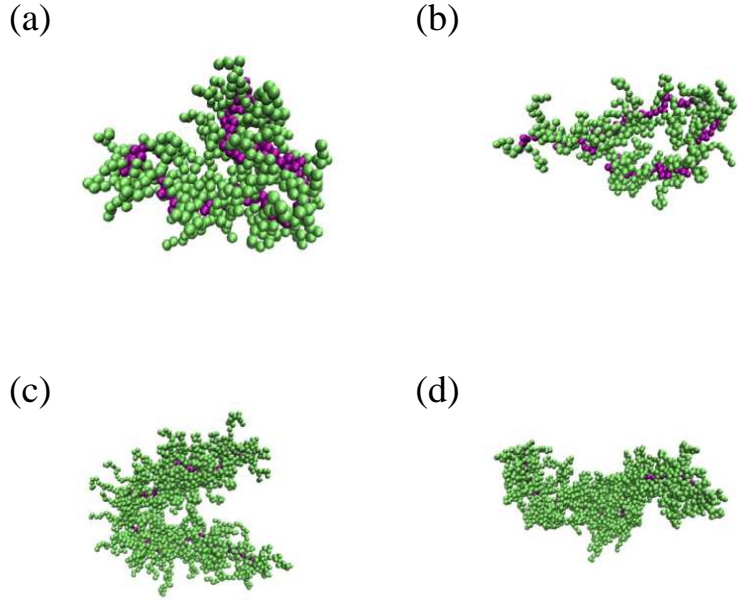} \hspace{0.8cm}
\caption{Selected snapshot pictures of equilibrated configurations of
bottle brush polymers. 
Backbone monomers (when visible) are displayed
in magenta color, side chain monomers in green. Cases shown refer to
$\sigma=0.5$, $N_b=100$, $N=10$, $T=3.0$ (a) and $T=4.0$ (b),
as well as $\sigma=1.0$, $N_b=100$, $N=20$, $T=3.0$ (c) and $T=4.0$ (d).}
\label{fig2}
\end{center}
\end{figure}

Fig.~\ref{fig2} shows a small selection of snapshot pictures of 
equilibrated bottle brush polymers. From these snapshot pictures it is
already clear that the side chains cause a significant stiffening of the
backbone, at least on a coarse-grained scale, and that bottle brushes where
$N_b$ is not very much larger than $N$ look like wormlike chains.
This conclusion corroborates pictures generated experimentally 
(by atomic force microscopy or electron microscopy techniques, 
e.g.~\cite{1,4,51}), but this observation should not mislead one to
claim that the Kratky-Porod wormlike chain model~\cite{52}
often employed to analyze such micrographs provides an accurate 
description of bottle brush polymers, as we 
shall see.

{
Of course, both the bead-spring model and the BFM are idealizations
of realistic comb-branched polymers, where also bond-angle
potentials and torsional potentials are present and contribute to the
local chain stiffness. Thus it is gratifying to note that nevertheless
measured structure factors of real bottle brush polymers~\cite{38} can be
mapped almost quantitatively to their simulated BFM counterparts~\cite{24},
if the lattice spacing is fixed at a length of a few Angstroms, and about 3
chemical monomers are mapped onto two effective monomers of the BFM.
Residual minor discrepancies may to some extent be due to solvent
quality effects~\cite{24}.}

\section{Side-chain and backbone linear dimensions and attempts
to extract ``the'' persistence length of bottle brush polymers}

Fig.~\ref{fig3} presents log-log plots of the normalized mean square 
gyration radius
of the side chains $\langle R_{gs}^2 \rangle/(l_b^2 N)$ as a 
function of side chain length $N$ ($l_b$ is the bond length
between successive monomers),
comparing data for grafting densities $\sigma=0.5$ and $\sigma=1.0$, 
and two temperatures 
($T=3.0$ and $T=4.0$, respectively). Three different backbone lengths
($N_b=50$, $100$, and $200$) are included, but the dependence of the 
data on backbone length is not visible on the scale of the graph.
Such a dependence on $N_b$ would be expected due to effects near the chain ends
of the backbone: in Ref.~\cite{24} it was shown that there is less stretching
of the side chains in the direction perpendicular to the backbone near the
chain ends of the backbone than in the central part of the backbone.
However, this effect is almost completely compensated by the fact that
side chains near the backbone chain ends are more strongly stretched in the
direction of the backbone (side chains then are oriented away from the backbone
chain ends). This fact has already been seen for bottle brushes with rigid
backbone but free ends~\cite{53}. 

We have also checked that for the bead spring model average side
chain linear dimensions for bottle brushes with rigid and flexible backbones
are identical for the bond fluctuation model, 
at least in the accessible parameter
regime. For the bond fluctuation model a similar equivalence has been
found also for the radial density profile of the monomers~\cite{54}.
Fig.~\ref{fig4} shows then the temperature dependence of 
$\langle R_{gs}^2 \rangle$ for the different choices of $N$,
demonstrating that the side chain extension is independent of $N_b$ and agrees
well between rigid and flexible backbones. The small deviations observable
between the rigid and the flexible backbone cases for large $N$ are due to
residual systematic errors of the simulations for the flexible backbone.
The present work therefore suggests that the equivalence between bottle brushes with
rigid and flexible backbones carries over to chains in
variable solvent quality, down to the Theta point (but we caution the 
reader that this equivalence will break down for poor solvents, in the regime
of intermediate grafting densities where for rigid backbones pearl-necklace
structures occur~\cite{29,30}). 

  The straight lines on the log-log plots in Fig.~\ref{fig3} illustrate 
the empirical power law
\begin{equation}
  \langle R_{gs}^2 \rangle \propto N^{2\nu_{\rm eff}}
\label{eq7}
\end{equation}
consistent with corresponding experiments (see~\cite{1,4} and references 
quoted therein) and we find that $\nu_{\rm eff}$ decreases with decreasing 
solvent quality and with decreasing grafting density. In the good 
solvent regime, $\nu_{\rm eff} > \nu=0.588$~\cite{55}, the established value
of the exponent describing the swelling of linear polymers under good solvent 
conditions, $R_g(N) \propto N^{\nu}$ for $N \rightarrow \infty$~\cite{8,9}.
As expected, $\nu_{\rm eff}$ is still distinctly smaller than the value
predicted from scaling~\cite{3}, 
{$\nu_{\rm eff}=2\nu/(1+\nu) \approx 0.74$}
or SCFT~\cite{19}, $3/4$.
{But the slight enhancement of $\nu_{\rm eff}$ with respect to
its value for free chains ($\nu=1/2$ at the Theta point, $T=3.0$;
$\nu=0.588$ for good solvents) is evidence that the side chains
interact with each other, which is a prerequisite for the expected
induced stiffening of the backbone chain.}

  We now turn to the backbone linear dimensions. Here we first caution the
reader that in good solvent conditions we clearly also expect that the 
mean square end-to-end distance of a bottle-brush polymer satisfies the standard
scaling relation
\begin{equation}
\langle R_{eb}^2 \rangle \propto N_b^{2\nu}\, , \qquad  \nu=0.588\, , \qquad
N_b \rightarrow \infty \, ,
\label{eq8}
\end{equation}
but the longer the side chain length the larger $N_b$ must be chosen such that
Eq.~(\ref{eq8}) can be verified. We start the discussion with data from the
bond fluctuation model, where (for $N\le 24$) data up to $N_b=1027$ are 
available~\cite{10}. Fig.~\ref{fig5} hence shows a log-log plot of 
$\langle R_{eb}^2 \rangle / (2 l_b^2 N_b^{2\nu}$) versus $N$. 
The factor $2$ in the denominator is arbitrarily chosen, since for 
Gaussian chains {(}$\nu=1/2${)} 
the result would simply be the persistence
length $l_p$ in units of the bond length~\cite{10}. Thus, it was
suggested~\cite{10} that in the excluded volume case one could introduce
an effective persistence length $l_{p,R}(N)$ via the definition
\begin{equation}
  l_{p,R} (N)= \langle R_{eb}^2 \rangle / (2 l_b N_b^{2\nu})\, ,
\qquad N_b \rightarrow \infty \, .
\label{eq9}
\end{equation}
Then Fig.~\ref{fig5} can be interpreted as a plot of $l_{p,R}(N)/l_b$
versus $N$ in the region where it is basically independent of $N_b$. The
quantity $l_{p,R}(N)$ defined in this way is  
compatible with an effective power law,
\begin{equation}
   l_{p,R}(N) \propto N^{\zeta}\, ,
\label{eq10}
\end{equation}
but the effective exponent $\zeta$ clearly depends on $N_b$, if $N_b$
is not chosen large enough (Fig.~\ref{fig6}). Thus, Fig.~\ref{fig5}
clearly shows that $l_{p,R}(N)$ is not a quantity characterizing the 
intrinsic stiffness of a bottle-brush polymer. For large $N_b$, the exponent
$\zeta$ seems to saturate 
at a value close to $0.65$ and $0.74$ for $\sigma=0.5$ and $\sigma=1.0$, 
respectively, but for chain lengths $N_b$ in the range 
from $N_b=67$ to $N_b=131$ it is only in the range from $\zeta=0.41$
to $\zeta=0.55$ and from $\zeta=0.48$ to $\zeta=0.67$ for $\sigma=0.5$
and $\sigma=1.0$, respectively.

   Thus, it is no surprise that for the bead-spring model choosing $N_b=50$,
$100$, or $200$ values of this effective exponent $\zeta$ in a similar range
are found (Fig.~\ref{fig7}). Of course, one must be aware that there is 
no strict one-to-one correspondence~\cite{56} between the meaning of chain 
lengths $N_b$ and $N$ in different models: actually there may be the
need for conversion factors $N_b^{\rm (BFM)}/N_b^{\rm (BS)}$ and 
$N^{\rm (BFM)}/N^{\rm (BS)}$ between the bead spring (BS) model and the bond
fluctuation model (BFM). The same fact is true when we compare simulations
to experiments; e.g., the data of Rathgeber et al.~\cite{38} could
be mapped to the BFM~\cite{24} implying an equivalence between 
$N_b^{\rm (exp)}=400$ and $N_b^{\rm (BFM)}=259$ and between $N^{\rm (exp)}=62$
and $N^{\rm (BFM)}=48$, for instance. We also note that with decreasing 
temperature $\langle R_{eb}^2 \rangle$ decreases, irrespective of
$N$ (and also $\zeta$ decreases).
{Again we remind the reader that this parameterization of 
our data in
terms of effective exponents (also used in related experimental
work~\cite{17,38} only can serve to indicate the place in an extended
crossover region to which the data belong.}

  When we analyze the gyration radius of the backbone as a function of backbone
chain length for the BS model (Figs.~\ref{fig8}) we find that over a restricted 
range of $N_b$ one can fit the data by effective exponents again, 
$\langle R_{gb}^2 \rangle^{1/2} \propto N_b^{\nu_{\rm eff}}$, 
with $0.55 \le \nu_{\rm eff} \le 0.95$, and it is seen that $\nu_{\rm eff}$
increases systematically with side chain length $N$, and 
$\nu_{\rm eff}$ for $T=4.0$ is larger than for $T=3.0$, 
for the same choice of $N$.
Of course, this variation of the effective exponent is just a reflection of
a gradual crossover from the rod-like regime 
($\langle R_g^2 \rangle^{1/2} \propto N_b$) to the self-avoiding 
walk-like behavior of swollen coils
($\langle R_g^2 \rangle^{1/2} \propto N_b^{\nu}$, with $\nu \approx 0.588$)
in the good solvent regime $(T=4.0)$ or Gaussian-like coils
($\langle R_g^2 \rangle^{1/2} \propto N_b^{1/2}$) for the Theta 
point ($T=3.0$), respectively. Similar plots as in Fig.~\ref{fig8} but for
the BFM are shown in Fig.~\ref{fig9}(a)(b). 
Due to the availability of equilibrated data for much longer backbone 
chain lengths $N_b$
for the BFM, the plot of $\langle R^2_{gb} \rangle / (2 l_b^2 N_b^{2\nu})$
versus $N_b$ (Fig.~\ref{fig9}(c)) indeed shows that the data settle down to a 
horizontal plateau implying that the asymptotic region where Eq.~(\ref{eq9})
can be applied indeed is reached. 
Instead of $R_{eb}^2$
in Eq.~(\ref{eq9}), we use $R_{gb}^2$ such that the estimate of the 
$N$-dependent effective 
persistence length $l_{p,Rg}$ is given by the horizontal line in  
Fig.\ref{fig9}(c).
Actually, for fixed side chain length $N$
the different choices of $N_b$ can be collapsed on a master curve, when
$N_b$ is rescaled by $s_{\rm blob}$~\cite{6}, which can be interpreted as 
an estimate for the number of segments per persistence length.
In our previous work~\cite{6,11}, we have shown that bottle brushes 
under very good solvent conditions can be described as self-avoiding walks
of effective ``blobs", having a diameter that is just twice the cross-sectional
radius of the bottle-brush, and containing a number $s_{\rm blob}$ of 
backbone monomers. Note that hence $s_{\rm blob}$ is not a fit parameter,
but has been determined independently.
This data collapse on a master curve is demonstrated for the BFM 
for the case $\sigma=1$, $N=6$, $12$, $18$,
and $24$ (Fig,~\ref{fig9}(d)). 

  Since ``the'' persistence length of polymers in the standard 
textbooks~\cite{7,8,9} is traditionally defined from the decay of bond
orientational correlations along the chains, we next focus on this 
quantity. Defining the bond vectors $\vec{a}_i$ in terms of the monomer
positions $\vec{r}_i$ as $\vec{a}_i=\vec{r}_{i+1}-\vec{r}_i$,
$i=1, \ldots,N_b-1$, this bond orientational correlation is defined as
\begin{equation}
   \langle \cos \theta(s) \rangle = l_b^{-2} \frac{1}{N_b-1-s}
\sum_{i=1}^{N_b-1-s} \langle \vec{a}_i \cdot \vec{a}_{i+s} \rangle
\label{eq11}
\end{equation}
Note that $\langle \vec{a}_i^2 \rangle = l_b^2$ and hence
$\langle \cos \theta(0) \rangle = 1$, of course. Considering the limit
$N_b \rightarrow \infty$, and assuming Gaussian chain statistics,
one obtains an exponential decay, since then $\langle \cos \theta(s) \rangle
=\langle \cos \theta (1) \rangle^s=\exp \left[ s \ln \langle \cos \theta
(1) \rangle \right]$, and thus
{\begin{equation}
  \langle \cos \theta (s) \rangle = \exp \left[ -s\ell_b/l_p \right]\, ,
\qquad l_p^{-1}=-\ln \langle \cos \theta (1) \rangle/\ell_b \, .
\label{eq12}
\end{equation}}
However, it has been shown by scaling and renormalization group
arguments~\cite{57} and verified by simulations~\cite{10,11} that in the good
solvent case there actually occurs a power law decay
\begin{equation}
   \langle \cos \theta (s) \rangle \propto s^{-\beta}\, , \qquad
\beta=2(1-\nu) \approx 0.824 \, , \qquad s \rightarrow \infty 
\label{eq13}
\end{equation}
while for chains at the Theta point~\cite{58} or in melts~\cite{59} one has
\begin{equation}
  \langle \cos \theta (s) \rangle \propto s^{-3/2} \, , \qquad 
 s  \rightarrow \infty \, .
\label{eq14}
\end{equation}
As far as bond orientational correlations are concerned, Gaussian chain 
statistics hence is misleading for polymers, under all circumstances.
However, Shirvanyants et al.~\cite{58} suggested that for semiflexible
polymers (at the Theta point) one can use still Eq.~(\ref{eq12})
but only for $1 \le s \le s^*$, where $s^* \propto l_p / l_b$
controls the crossover from the simple exponential decay, Eq.~(\ref{eq12}),
to the power law, Eq.~(\ref{eq14}). Indeed, for a simple self-avoiding
walks (SAW) model on the simple cubic lattice, where chain stiffness was
controlled by an energy cost $\varepsilon_b$ when the chain makes a $90^o$
kink on the lattice~\cite{11}, it was shown that this suggestion~\cite{58}
works qualitatively, also in the good solvent case, with 
$l_p \propto \exp(\varepsilon_b / k_BT)$ for $\varepsilon_b/k_BT \gg 1$,
although the crossover between Eqs.~(\ref{eq12}) and (\ref{eq13}) is not
sharp but rather spread out over a decade in the variable $s$.

Motivated by this finding~\cite{11}, Figs.~\ref{fig10}, \ref{fig11} hence 
present a few examples where
$\langle \cos \theta (s) \rangle$ is plotted vs. $s$ on a semi-log plot,
to test for a possible applicability of Eq.~(\ref{eq12}) for not too
large $s$. However, we find that in fact Eq.~(\ref{eq12}) does NEVER
hold for small $s$ ($s=1$, $2$, $3$), unlike the semiflexible SAW model
of Ref.~\cite{11}, rather there occurs a very fast decay of 
$\langle \cos \theta (s) \rangle$ following a strongly bent curve (only
two successive values of $s$ can always be fit to a straight line, of 
course, but there is never an extended regime where a straight line
through the origin, $\langle \cos \theta (s=0) \rangle=1$, would be
compatible with the data). It is also remarkable that this initial behavior
is almost independent of $\sigma$ and $N$; an obvious interpretation
is that the stiffening of the backbone caused by the presence of long
side chains is effective only on mesoscopic length scales along the backbone,
but not on the scale of a few subsequent backbone bonds, which maintain a 
high local flexibility. Only for $s \ge 4$ the data are compatible with
a relation
\begin{equation}
   \langle \cos \theta(s) \rangle = a \exp(-bs)\, , \qquad 
4\le s \le s_{\rm max}\, ,
\label{eq15}
\end{equation}
where $a$ and $b$ are phenomenological constants, and $s_{\rm max}$ depends on
both $N$, $\sigma$, and $T$ distinctly (but cannot be accurately obtained
from our simulations, 
because for $\langle \cos \theta (s) \rangle \le 0.03$ the statistical
accuracy of the data deteriorates.) Obviously, the relation
{(}Eq.~(\ref{eq12}){)} 
{$l_p^{-1} = -\ln \langle \cos \theta (1) \rangle/\ell_b$}
fails, but it seems tempting to identify an effective persistence length
$l_p^{\rm eff}$ as {$l_p^{\rm eff}/\ell_b=b^{-1}$}. 
However, when we would 
define the persistence length in this way, we obtain the result that 
$l_p^{\rm eff}$ depends on both $N_b$ and on $T$, not only on the
side chain length $N$ and grafting density $\sigma$ (Fig.~\ref{fig12} and 
Table~\ref{table1}).
{For the bead-spring model, one often estimates that the
length unit $(\sigma_{\rm LJ}=1)$ physically corresponds to about
$0.5\;{\rm nm}$. The data for $\ell_p^{\rm eff}(T)$ at $\sigma=1$ and
good solvent conditions for $N_b = 200$ then would span a range
from about $5$ to about $40\;{\rm nm}$, i.e. a similar range as
proposed in recent experiments$^{17}$. But}
already our previous work on the athermal bond fluctuation 
model~\cite{6,10,24} has given some evidence, that defining a persistence
length from a fit of the data for $\langle \cos \theta (s) \rangle$
to Eq.~(\ref{eq15}) is not suitable to obtain a measure of the local
intrinsic stiffness, since $l_p^{\rm eff}$ depends on $N_b$.
The present data show that $l_p^{\rm eff}$ depends on $T$ as well,
and hence is not just controlled by the chemical architecture of the
bottle brush (via the parameters, $\sigma$, $N$, and $N_b$),
but depends on solvent conditions as well.

  Thus, we argue that the physical significance of a persistence length
$l_p^{\rm eff}$ extracted from bond orientational correlations in
this way is very doubtful, even at the Theta point. We also note that
sometimes, due to curvature on the semi-log plot 
{(}e.g. Fig.~\ref{fig11}(d){)}
such fits are ill-defined.
 
Finally, we consider the possible validity of 
the Kratky-Porod result
for the end-to-end distance of the chains at the Theta temperature, where 
for $N_b \rightarrow \infty$ we have (apart from logarithmic corrections)
$\langle R_e^2 \rangle \propto N_b$, i.e. for this property
a formula analogous to Gaussian
chains holds. The Kratky-Porod result describes the crossover from rods
to Gaussian chains,
\begin{equation}
  \frac{\langle R_e^2 \rangle }{ 2 l_p L} = 1-\frac{l_p}{L}
\left[ 1-\exp(-L/l_p)\right] \, .
\label{eq16}
\end{equation}
We can define an effective exponent $\nu_{\rm eff}(L/l_p)$ in terms of the
logarithmic derivative of this function, Fig.~\ref{fig13} (a),(b),
\begin{equation}
   \frac{d\ln\left[\langle R_e^2 \rangle / (2 l_p L) \right]}
{d \ln (L/l_p)}=2\nu_{\rm eff}(L/l_p)-1 \, .
\label{eq17}
\end{equation}
Using the data for $\nu_{\rm eff}$ at $N_b=100$ from Fig.~\ref{fig8}(a)(c)
(and similar data not shown for $\langle R_{eb}^2 \rangle$ ) we hence obtain
estimates for the ratios $L / l_p$ for the various choices of $N$ for
both $\sigma=0.5$ and $\sigma=1$.
Note that the errors in our estimation of $\nu_{\rm eff}$ translate into
rather large errors in our estimates for $L/l_p$ (both these errors 
of our data are indicated in Fig.~\ref{fig13}(a)(b)). Since we expect that
the actual variation of $\langle R_{eb}^2 \rangle$ with $N_b$ exhibits slight
curvature (although this is hardly detected in Fig.~\ref{fig8}(a)(c)), we
use only this intermediate value of $N_b=100$ to estimate the relation between
$N$ and $L/l_p$, and we use neither $N_b=50$ nor $N_b=200$ for 
this purpose.

  If the contour length $L$ of the Kratky-Porod model, Eq.~(\ref{eq16}), would
simply be the ``chemical'' contour length $L_{\rm ch}=N_b l_b$, where the
bond length $l_b$ connecting two neighboring monomers along the backbone
is $l_b \approx 1$, our results for $L/l_p$ would readily yield explicit
results for $l_p$. However, using then Eq.~(\ref{eq16}) to compute 
{$\langle R_{eb}^2 \rangle/L^2$} yields an overestimation 
by about a factor of five
for $\sigma=1$.
This discrepancy proves that $L$ is significantly smaller than $L_{\rm ch}$.
{Since Eq.~(17) implies that there is a unique correspondence
between $\nu_{\rm eff}$ and the ratio $\ell_p/L$, the factor of
five discrepancy for $\langle R_e^2 \rangle$ (which is proportional
to the product of $\ell_p$ and $L$) means that both $\ell_p$ and $L$
must be smaller by the same factor (at about $\sqrt{5}$).}
Thus we define $L=N_b l_b^{\rm eff}$, where $l_b^{\rm eff}(<1)$
is interpreted as the average projection of a backbone bond on the direction
of the coarse-grained contour of the wormlike chain. Thus we also can use
the data shown in Fig.~\ref{fig13}(a)(b) to obtain explicit estimates
for $l_p(N)$, taking into account $L=N_b l_b^{\rm eff}$ instead
of $L=N_b l_b$. Interestingly, the mechanisms leading to
$l_b^{\rm eff}<1$ is also evident in the presence of the constant $a<1$ in
the fits of $\langle \cos \theta(s) \rangle=a \exp(-bs)$ in 
Figs.~\ref{fig10}, \ref{fig11}. 
Using then the estimates $l_b^{\rm eff}(\sigma=0.5) \approx 0.33$ and
$l_b^{\rm eff}$ ($\sigma=1) \approx 0.45$, the data for 
$\langle R_{eb}^2 \rangle / (2 l_p(N) L)$ at $T=3.0$ are roughly
compatible with Eq.~(\ref{eq16}), when we use $L/l_p(N)$ from 
Fig.~\ref{fig13}(a)(b) and take $L=l_b^{\rm eff}N_b$, as is shown in 
Fig.~\ref{fig13}(c)(d). 
In this way, we have defined a correction factor $a_r=\left(l_b^{\rm eff}
\right)^{-2}$ in Fig.~\ref{fig13}, which was assumed to depend neither on $N$
nor on $N_b$. Note that while only $N_b=100$ was used in Fig.~\ref{fig13}(a)(b),
data for $N_b=50$, and $200$ are included in Fig.~\ref{fig13}(c)(d).
The resulting values of $l_p(N)$ 
{which hence by construction do not depend on $N_b$}
(Table~\ref{table1}) are considered 
to be the most reliable estimates for the considered model.
However, we emphasize that Eq.~(\ref{eq16}) for bottle brushes is useful only
if the solution is at Theta conditions, but not in the good solvent regime.
Using the data for $\nu_{\rm eff}$ at $N_b=100$ from Fig.~\ref{fig8}(a)(c)
(and similar data not shown for $\langle R_{eb}^2 \rangle$ ) we hence obtain
estimates for the ratios $L / l_p$ for the various choices of $N$ for
both $\sigma=0.5$ and $\sigma=1$.

\section{Conclusions}
In the present paper, a coarse-grained bead-spring model for bottle
brush polymers was studied by Molecular Dynamics methods, varying both 
the chain length $N_b$ of the backbone and of the side chains ($N$),
for two values of the grafting density, under variable solvent conditions.
The main emphasis of the present work was a study of the various 
characteristic lengths describing the conformation of the macromolecule,
contrasting the behavior under Theta conditions with the behavior in
the good solvent regime. Also a comparison with corresponding results
for the bond fluctuation model has been performed; this athermal model
represents very good solvent conditions, and while for corresponding parameters
it yields results that are rather similar to the results for the bead
spring model in the good solvent regime, it has the advantage that much
larger values of $N_b$ can be studied.

\begin{figure}
\begin{center}
(a)\includegraphics[scale=0.28,angle=270]{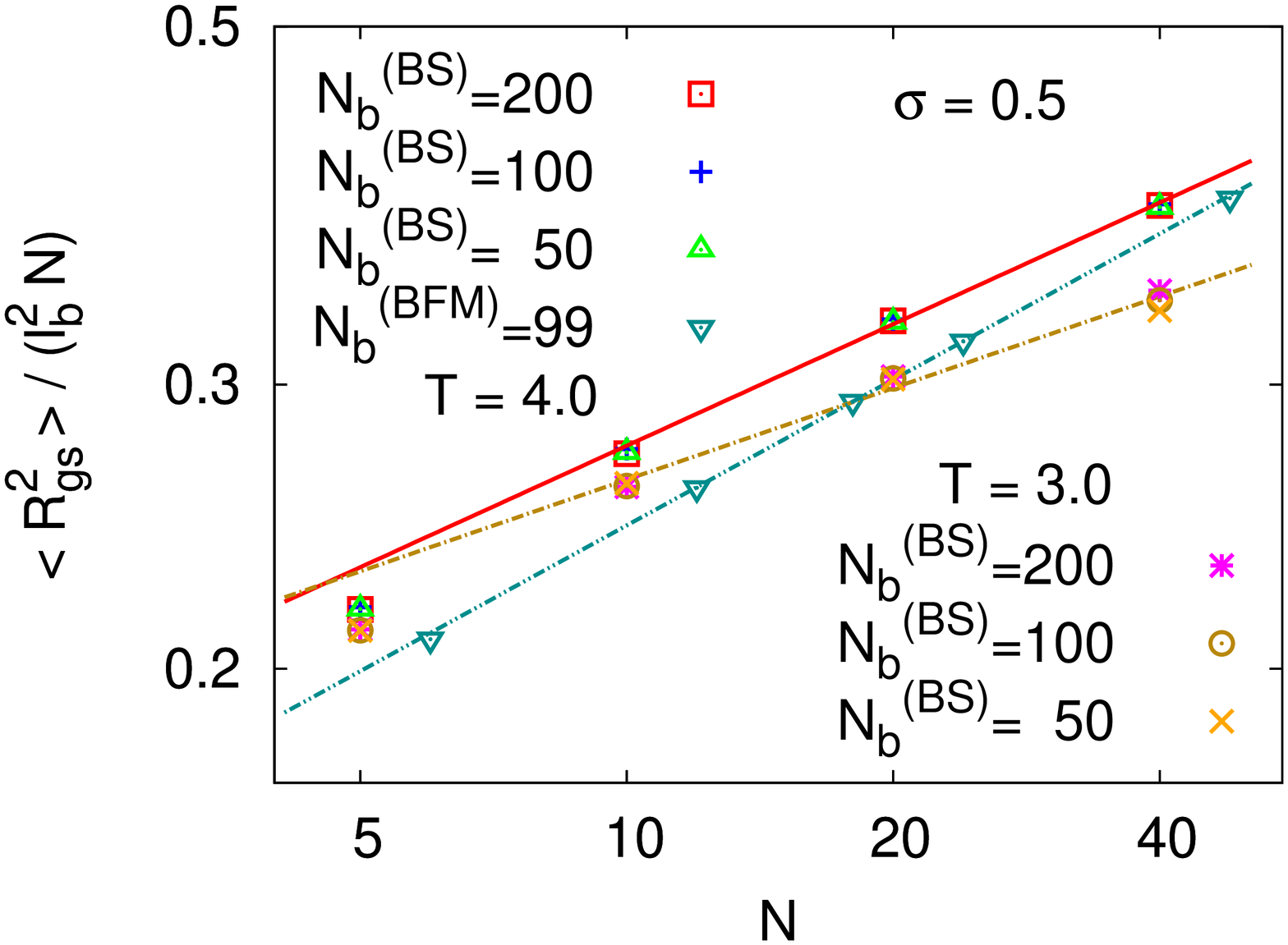}\hspace{0.8truecm}
(b)\includegraphics[scale=0.28,angle=270]{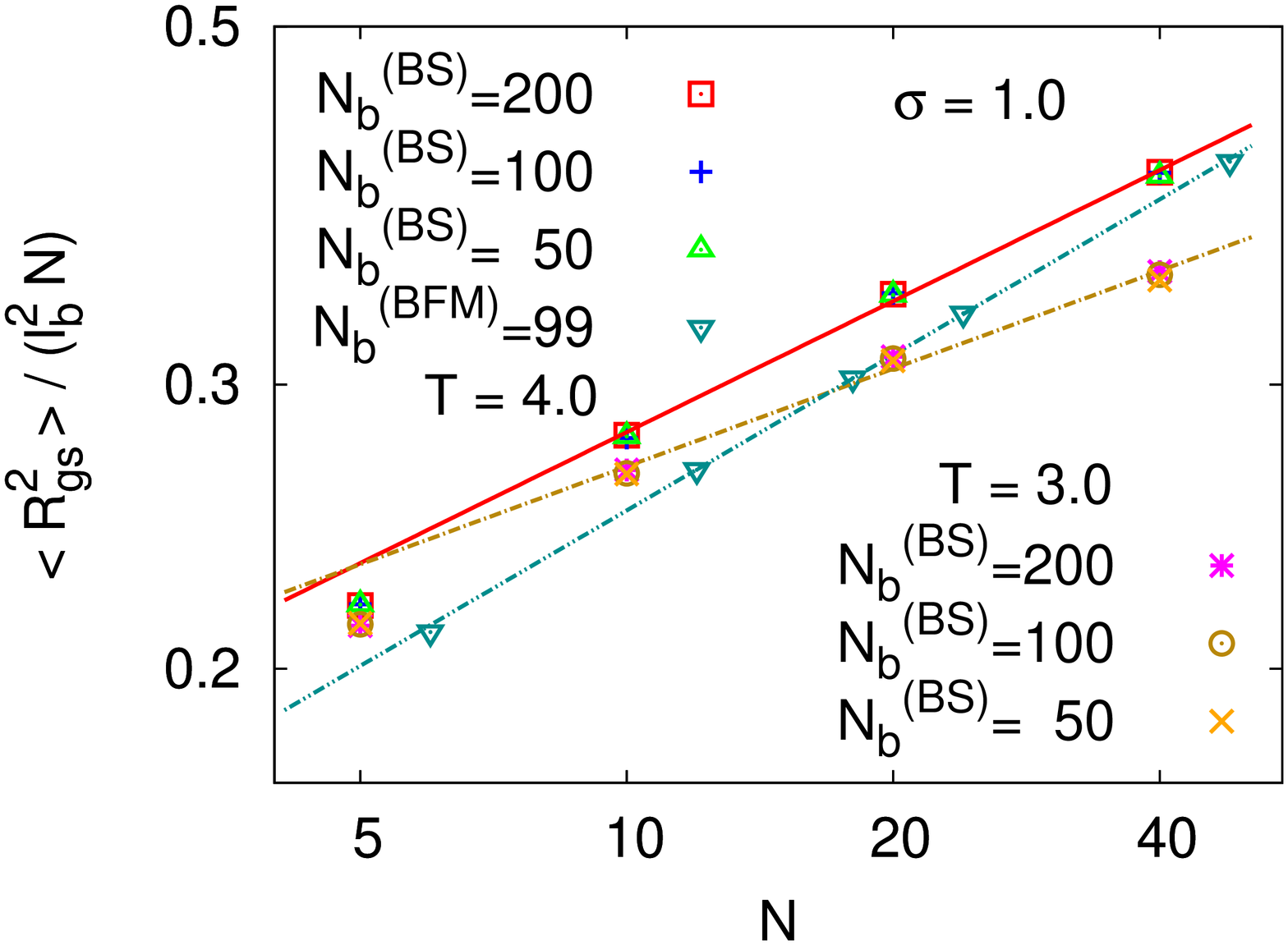}\\
\caption{Log-log plot of mean square radius of gyration of the side chains
normalized by their chain length and by the square bond length 
$l_b^2$ between
successive monomers,
$\langle R_{gs}^2 \rangle/(l_b^2N)$, versus side chain length $N$, 
for $\sigma=0.5$ (a)
and $\sigma=1.0$ (b). Data for the bead spring model at 
two temperatures ($T=3.0$ and $T=4.0$), as 
indicated in the figure are included, as well as three backbone chain lengths
($N_b=50$, $100$, and $200$, respectively). For comparison, also data for
the athermal bond fluctuation model 
{for comparable backbone chain length} are included.
Straight lines indicate effective exponents $\nu_{\rm eff} \approx0.60$ 
($T=3.0$)
or $\nu_{\rm eff} \approx 0.63$ ($T=4.0$) in case (a), and 
$\nu_{\rm eff} \approx 0.60$
($T=3.0$) or $\nu_{\rm eff}\approx 0.64$ ($T=4.0$) in case (b). For the
bond fluctuation model under very good solvent conditions 
the slightly larger effective exponent 
($\nu_{\rm eff} \approx 0.65$ for $\sigma=0.5$, and 
$\nu_{\rm eff} \approx 0.66$ for $\sigma=1.0$) than for the off-lattice 
model with $T=4.0$ results.}
\label{fig3}
\end{center}
\end{figure}

\begin{figure}
\begin{center}
\includegraphics[scale=0.28,angle=270]{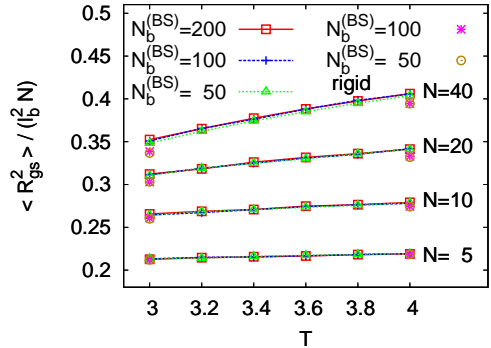}
\caption{Temperature dependence of the normalized mean square radius of gyration
$\langle R_{gs}^2 \rangle/ (l_b^2 N)$ for four chain lengths 
($N=5$, $10$, $20$, and $40$), and 
three backbone lengths ($N_b=50$, $100$, and $200$), respectively.
All data are for the bead spring model and $\sigma=1$.
Data for bottle-brush polymers with flexible backbone are connected by 
curves to guide the reader's eyes. Data are only shown for $T=3.0$ and
$4.0$ for the rigid backbone case.}
\label{fig4}
\end{center}
\end{figure}

   Among the quantities that have been studied are the end-to-end distances
of side chains and of the backbone, as well as the respective gyration
radii, but also mesoscopic lengths that are particularly popular when a
description in terms of the Kratky-Porod wormlike chain model is attempted,
namely ``the'' persistence length and ``the'' contour length of the 
wormlike chain. We have shown that for the range of rather short side 
chain lengths $N$ that is only accessible, either in simulation or experiment,
one often finds that a description in terms of power laws with effective
exponents is adequate. Specifically, the side chain radii vary as 
$\langle R_{gs}^2 \rangle^{1/2} \propto N^{\nu_{\rm eff}}$, and 
$\nu_{\rm eff} \approx 0.60$ (at the Theta temperature $T=T_\theta$)
or $\nu_{\rm eff}\approx 0.63$ to $0.66$ (in the good solvent regime),
see Fig.~\ref{fig3}. These effective exponents (for the present models with 
flexible backbones) are the same as for otherwise equivalent models with
stiff backbones, and also are practically independent of the backbone
chain length. While the resulting effective exponents are systematically
smaller than the asymptotic values predicted theoretically for the limit
$N \rightarrow \infty$, $\nu\approx 0.74$ (good solvents) or $\nu=2/3$
(Theta solvents), respectively, they are comparable to the results
of pertinent experiments: Zhang et al.~\cite{1,60} report 
$\nu_{\rm eff} \approx 0.67$ for a bottle brush where the chemical nature of
the main chain and the side chains is identical. For other systems exponents
in the range $0.56 \le \nu_{\rm eff} \le 0.67$ are reported~\cite{17},
again values comparable to our findings. Some experimental studies also
report that somewhat different results occur when different solvents 
are used~\cite{1,4,12,17,18}, but a systematic variation of solvent
quality to our knowledge has not yet been performed. Our results have shown
that for $T \ge T_\theta$ the side chain radii vary only very weakly with
temperature (which in our model causes the change in solvent quality),
see Fig.~\ref{fig4}.

\begin{figure}
\begin{center}
(a)\includegraphics[scale=0.28,angle=270]{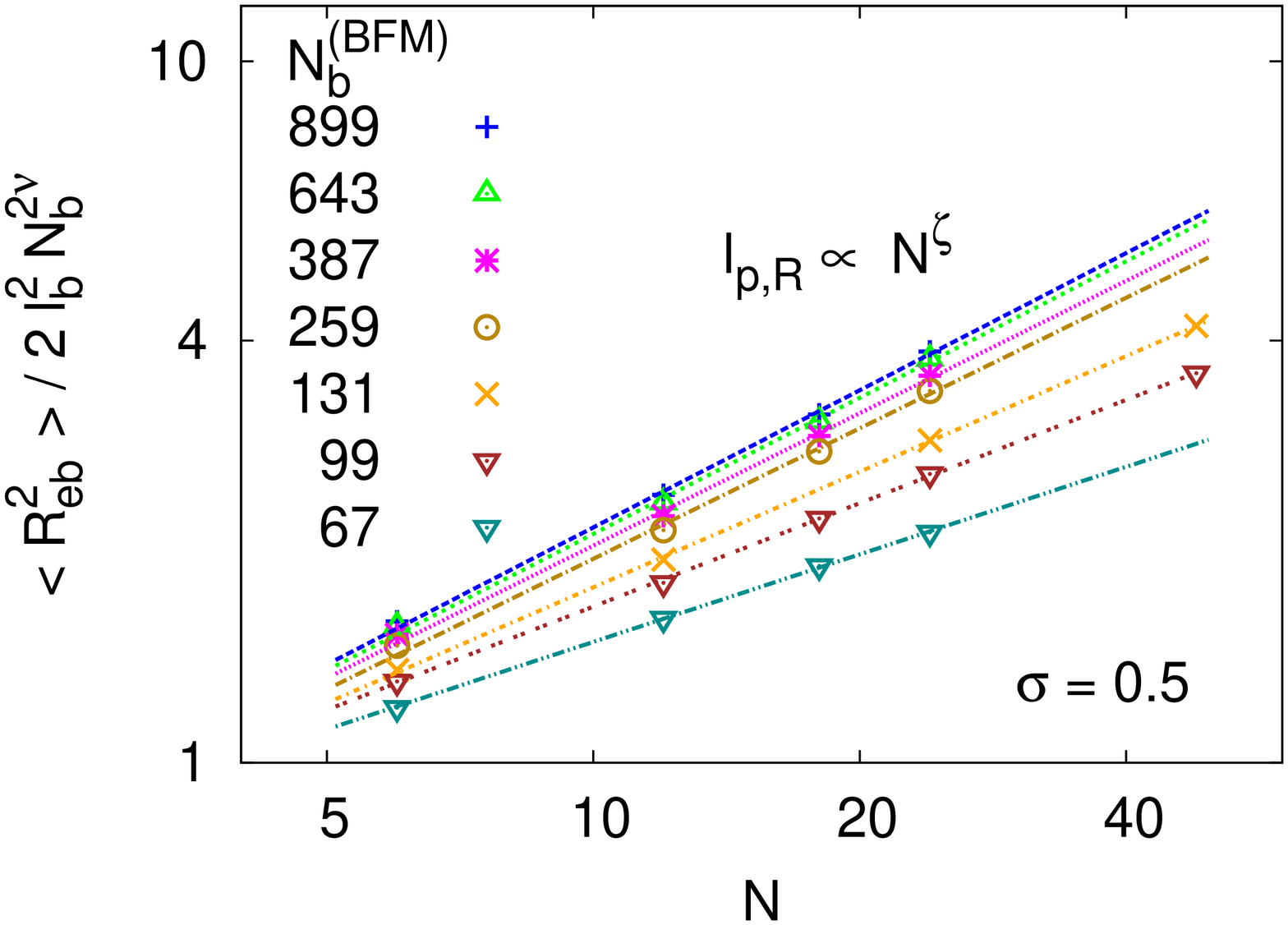}\hspace{0.8truecm}
(b)\includegraphics[scale=0.28,angle=270]{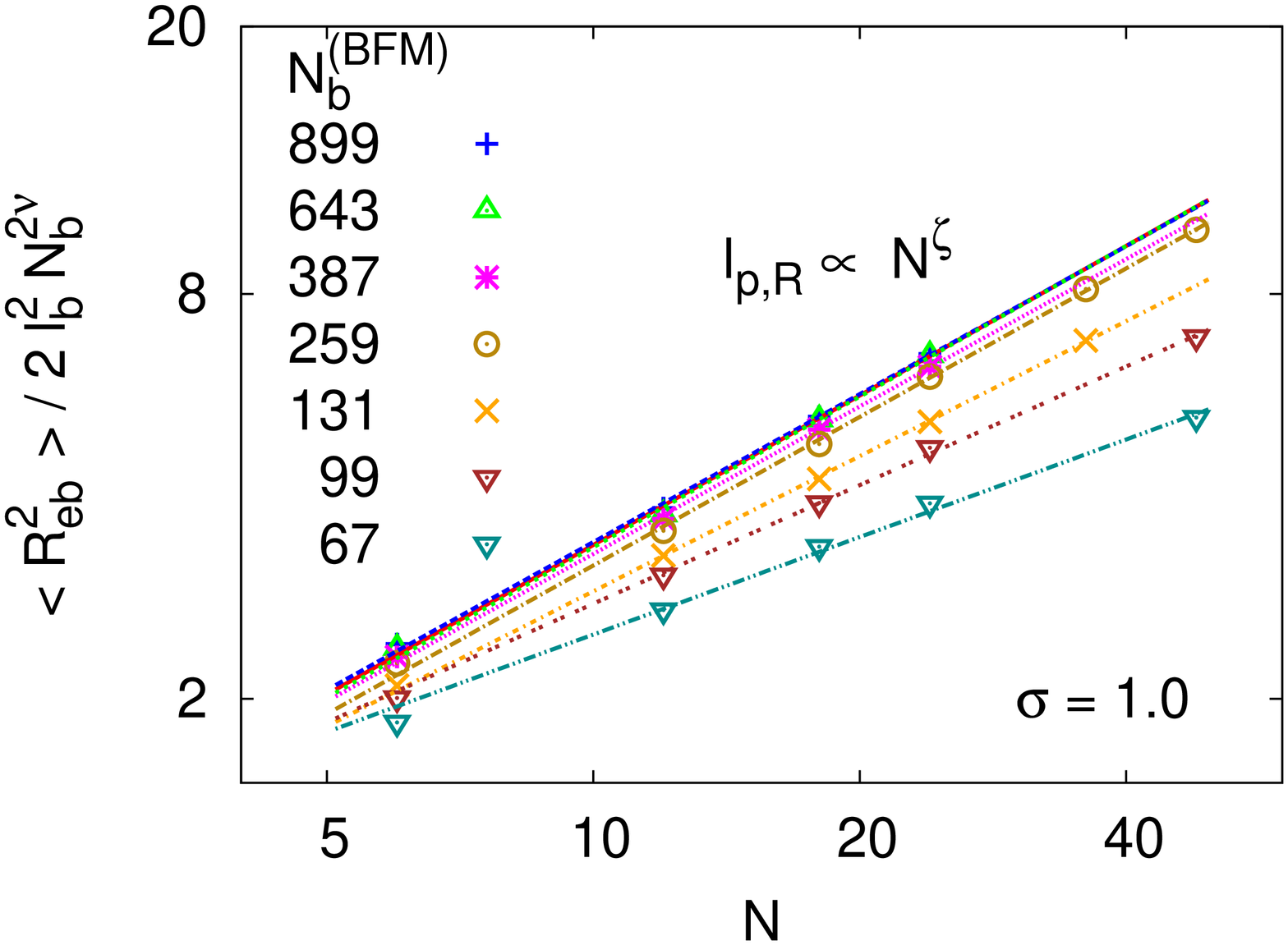}\\
\caption{Log-log plot of {$\langle R_{eb}^2 \rangle / (2 l_b^2 N_b^{2\nu})=\ell_{p,R}/\ell_b$}
versus side chain length $N$, for the bond fluctuation model with 
$\sigma=0.5$ (a) and $\sigma=1.0$ (b)
including different choices of $N_b$, as indicated.}
\label{fig5}
\end{center}
\end{figure}

\begin{figure}
\begin{center}
(a)\includegraphics[scale=0.28,angle=270]{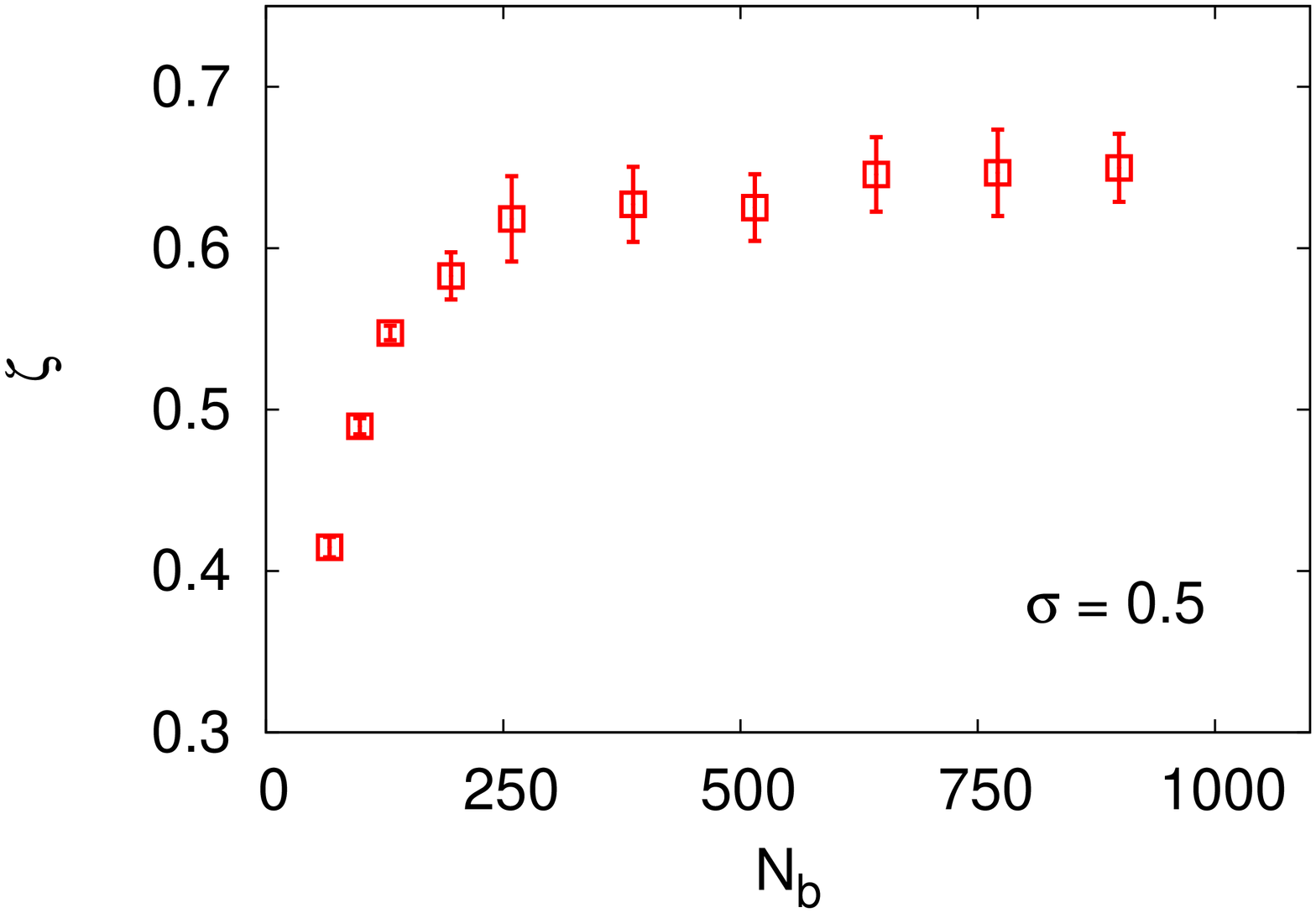}\hspace{0.8truecm}
(b)\includegraphics[scale=0.28,angle=270]{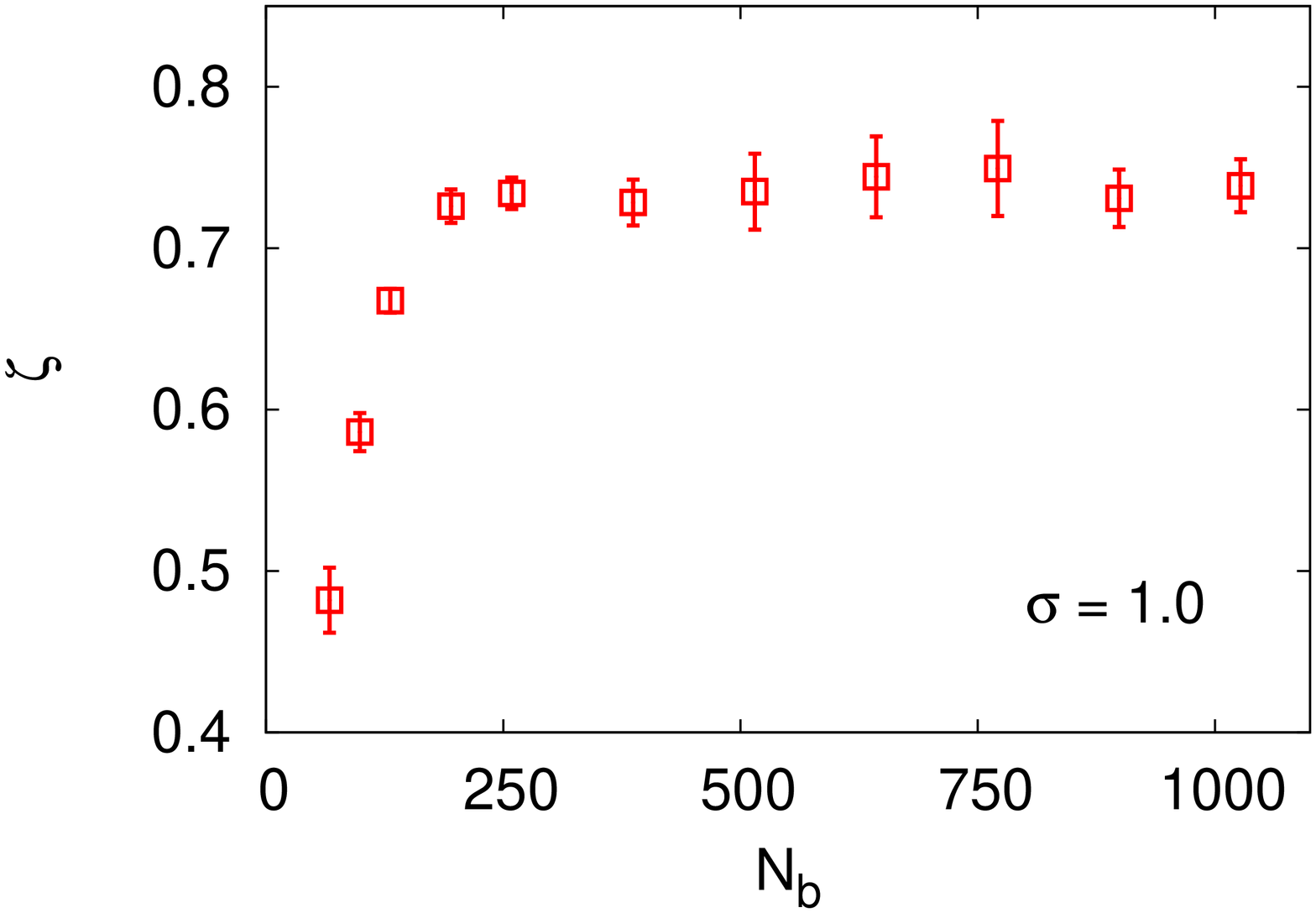}\hspace{0.8truecm}
\caption{Plot of the effective exponent $\zeta$, extracted from the
slope of the straight lines in Fig.~\ref{fig5}, versus the backbone
chain length $N_b$ for $\sigma=0.5$ (a) and $\sigma=1.0$. 
All data are for the athermal bond fluctuation model.}
\label{fig6}
\end{center}
\end{figure}

\begin{figure}
\begin{center}
(a)\includegraphics[scale=0.28,angle=270]{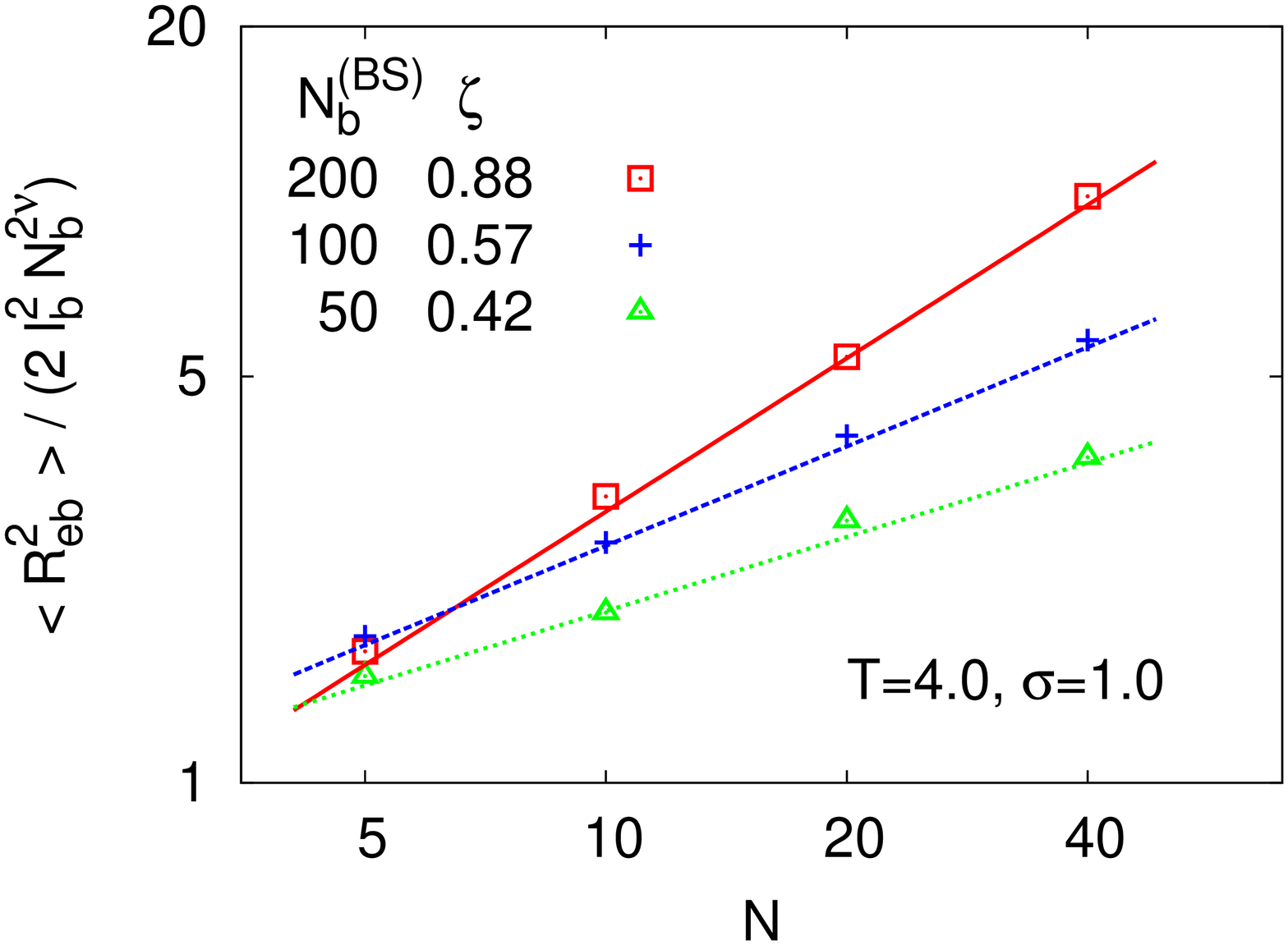}\hspace{0.8truecm}
(b)\includegraphics[scale=0.28,angle=270]{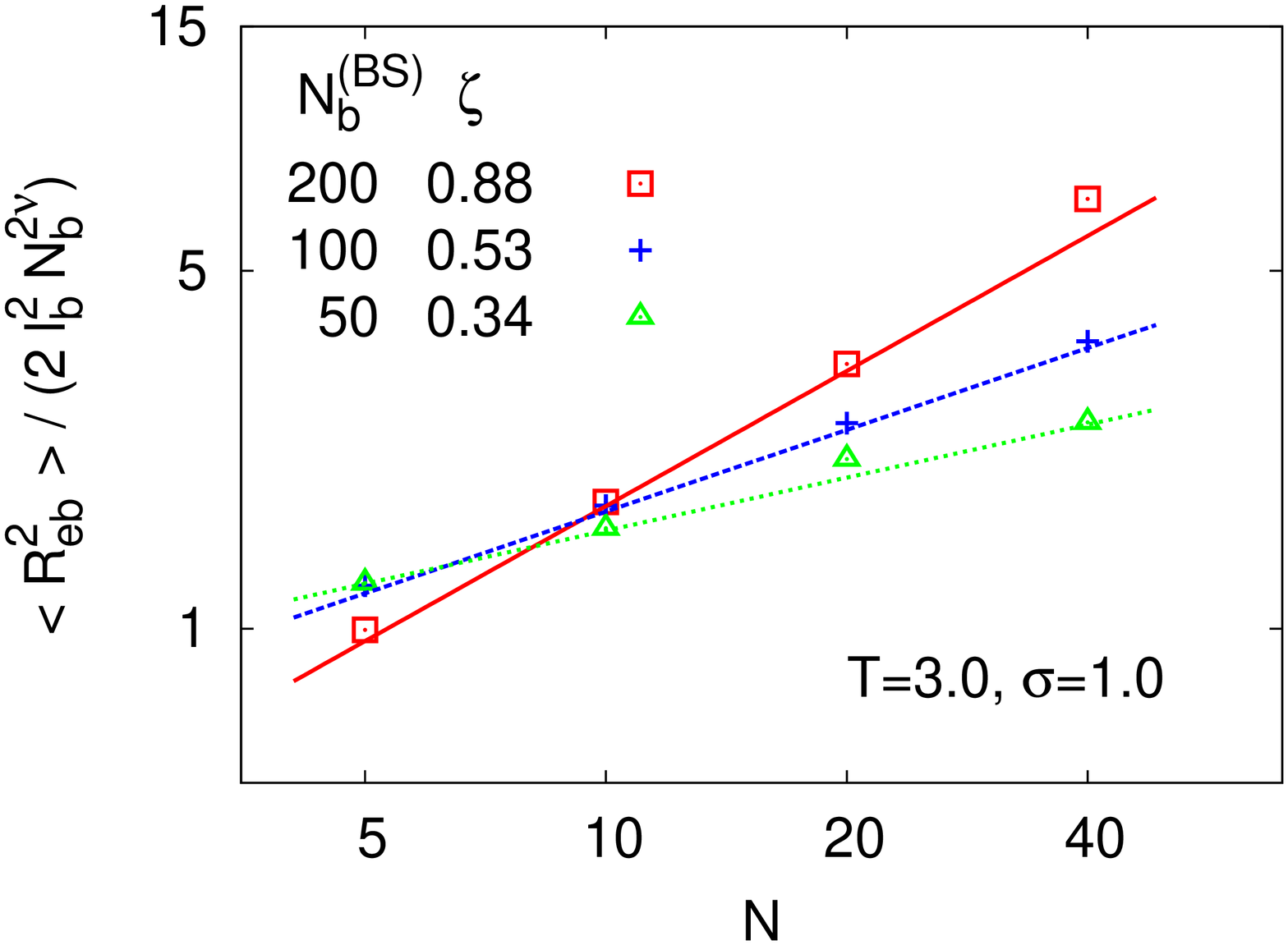}\\
\caption{Log-log plot of {$\langle R_{eb}^2 \rangle/(2l_b^2 N_b^{2\nu})=\ell_{p,R}/\ell_b$}
versus $N$. All data are for the bead spring model with $\sigma=1.0$
at both $T=4.0$ (a) and $T=3.0$ (b). Three backbone lengths are shown as
indicated.} 
\label{fig7}
\end{center}
\end{figure}

  Also the backbone end-to-end distance shows a power law variation
with the side chain length, $\langle R_{eb}^2 \rangle \propto N^\zeta$
(Fig.~\ref{fig5} - \ref{fig7}), reflecting thus the systematic backbone 
stiffening with increasing side chain length. In this case, however,
the effective exponent depends on the backbone chain length $N_b$,
for $N_b\le 300$, before a saturation at $\zeta\approx 0.65$
($\sigma=0.5$) or $0.75$ ($\sigma=1$) occurs, in the case of the 
bond fluctuation model. Even somewhat larger exponents are observed for
the bead-spring model (up to $\zeta \approx 0.88$, Fig.~\ref{fig7}).

\begin{figure}
\begin{center}
(a)\includegraphics[scale=0.28,angle=270]{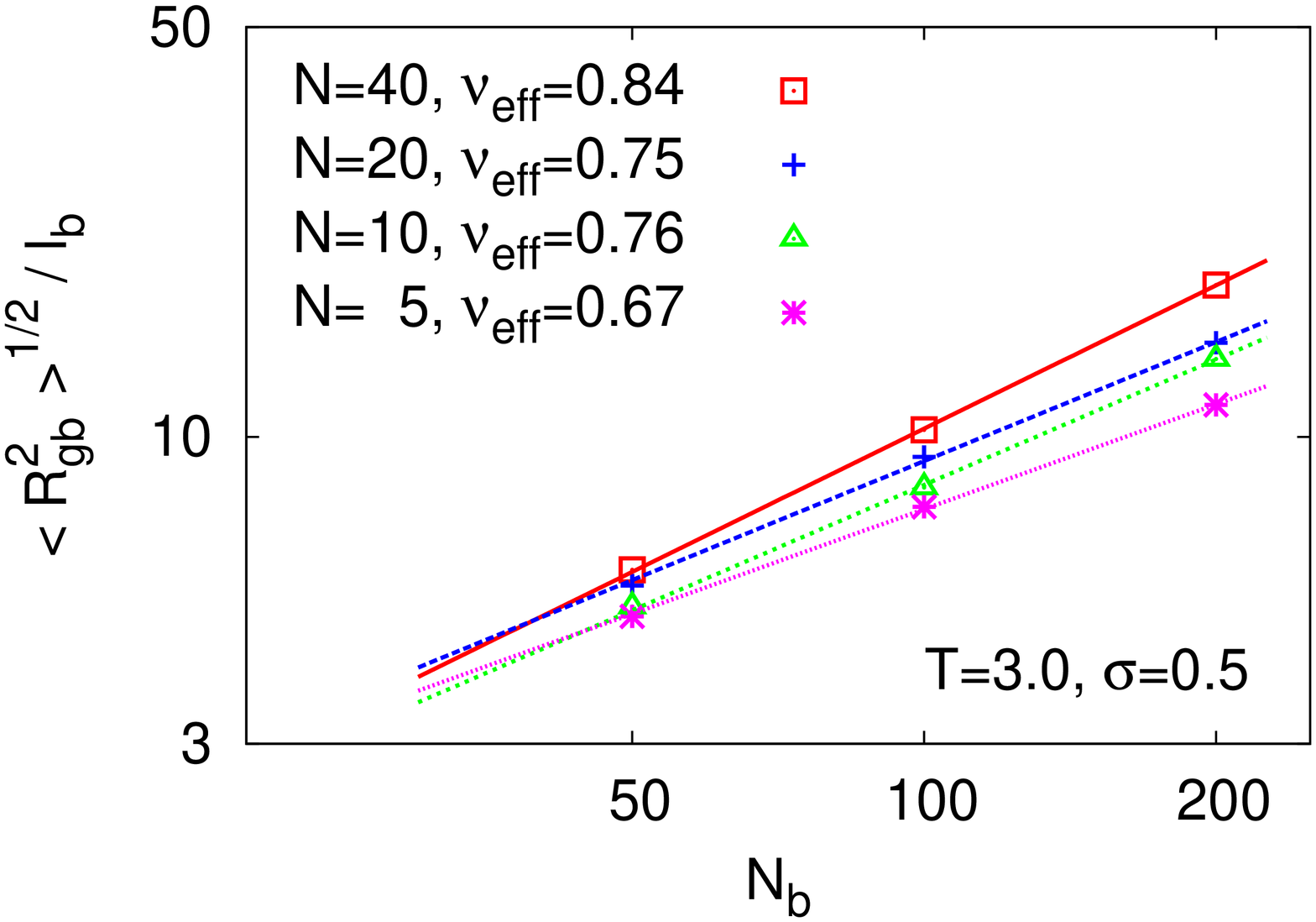}\hspace{0.8truecm}
(b)\includegraphics[scale=0.28,angle=270]{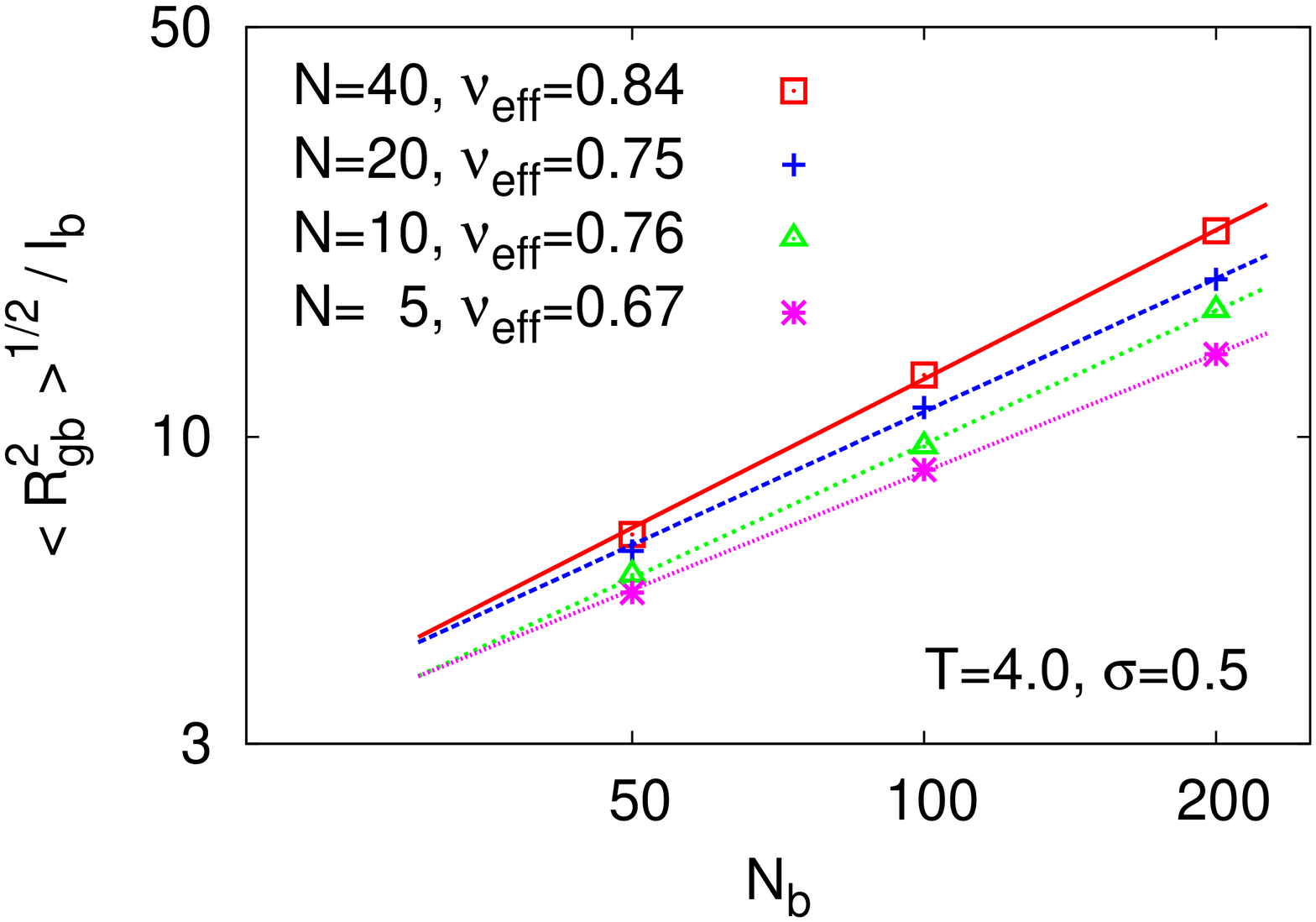}\\
(c)\includegraphics[scale=0.28,angle=270]{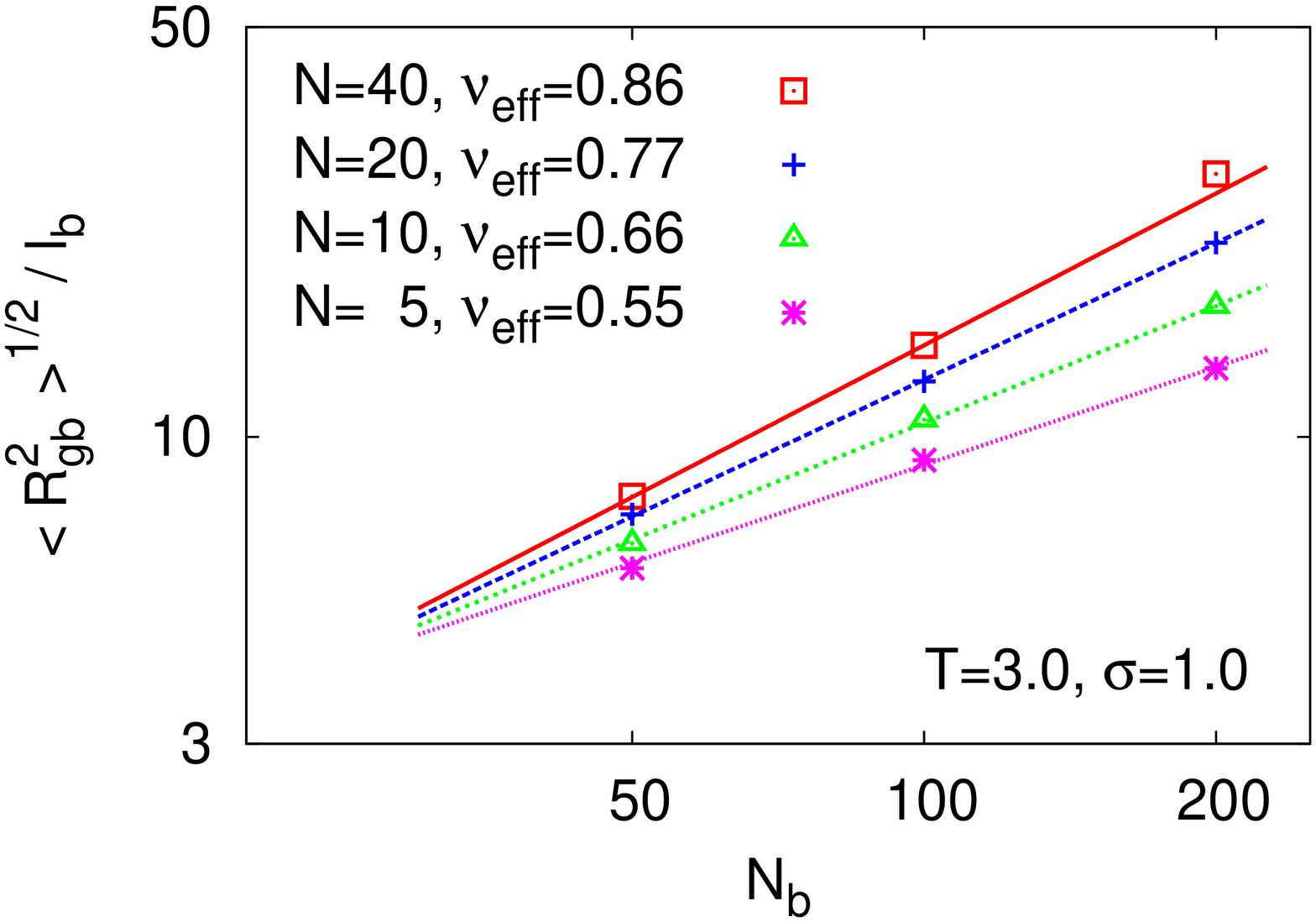}\hspace{0.8truecm}
(d)\includegraphics[scale=0.28,angle=270]{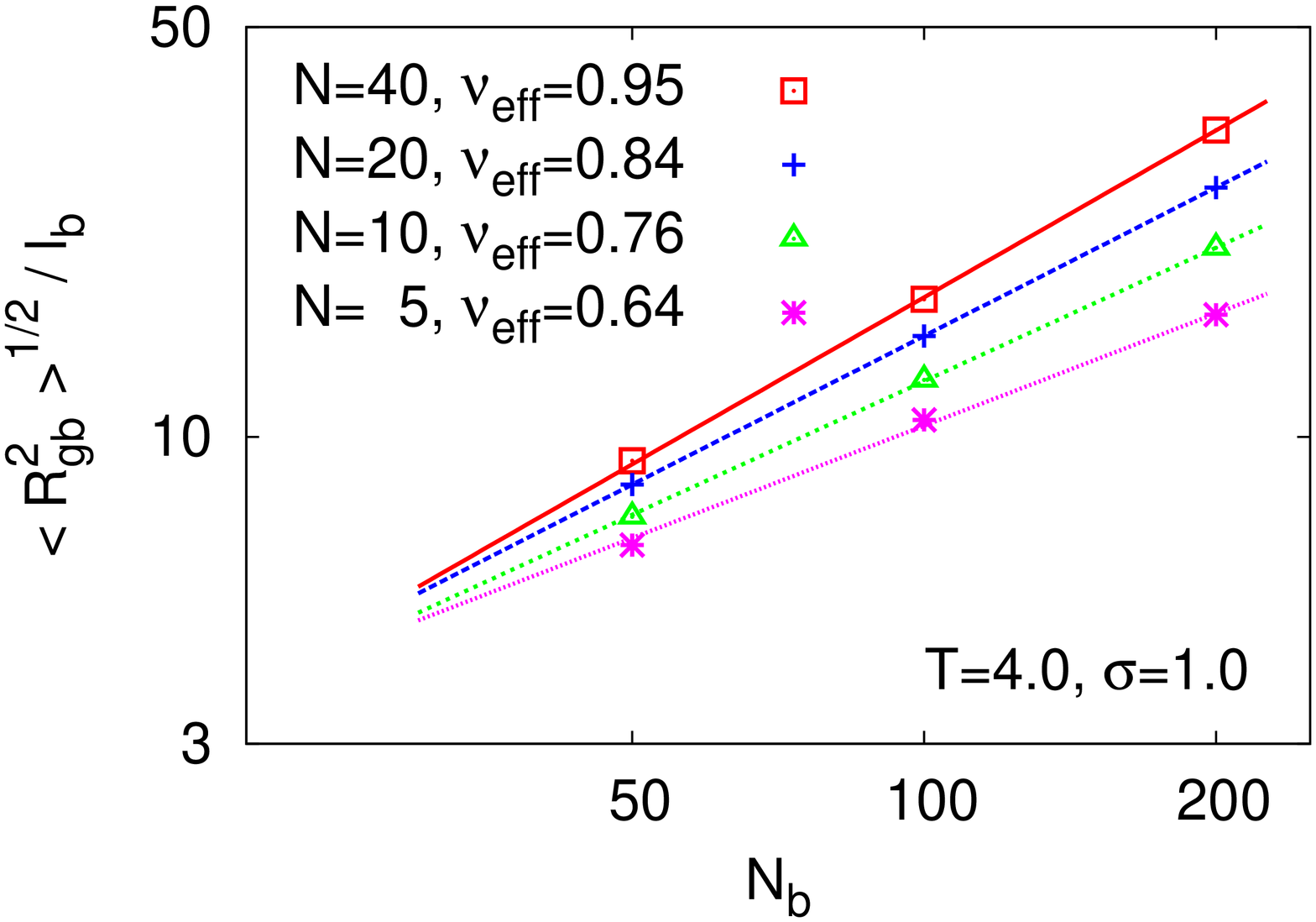} \\
\caption{Log-log plot of 
{$\langle R_{gb}^2 \rangle^{1/2}/\ell_b$}
versus backbone chain length $N_b$, for $\sigma=0.5$ (a)(b) and
$\sigma=1.0$ (c)(d), including both $T=3.0$ (a)(c)
and $T=4.0$ (b)(d), for the bead-spring model. Several side chain
lengths $N$ are included as indicated. Straight lines indicate a fit
with effective exponents $\nu_{\rm eff}$.}
\label{fig8}
\end{center}
\end{figure}

\begin{figure}
\begin{center}
(a)\includegraphics[scale=0.28,angle=270]{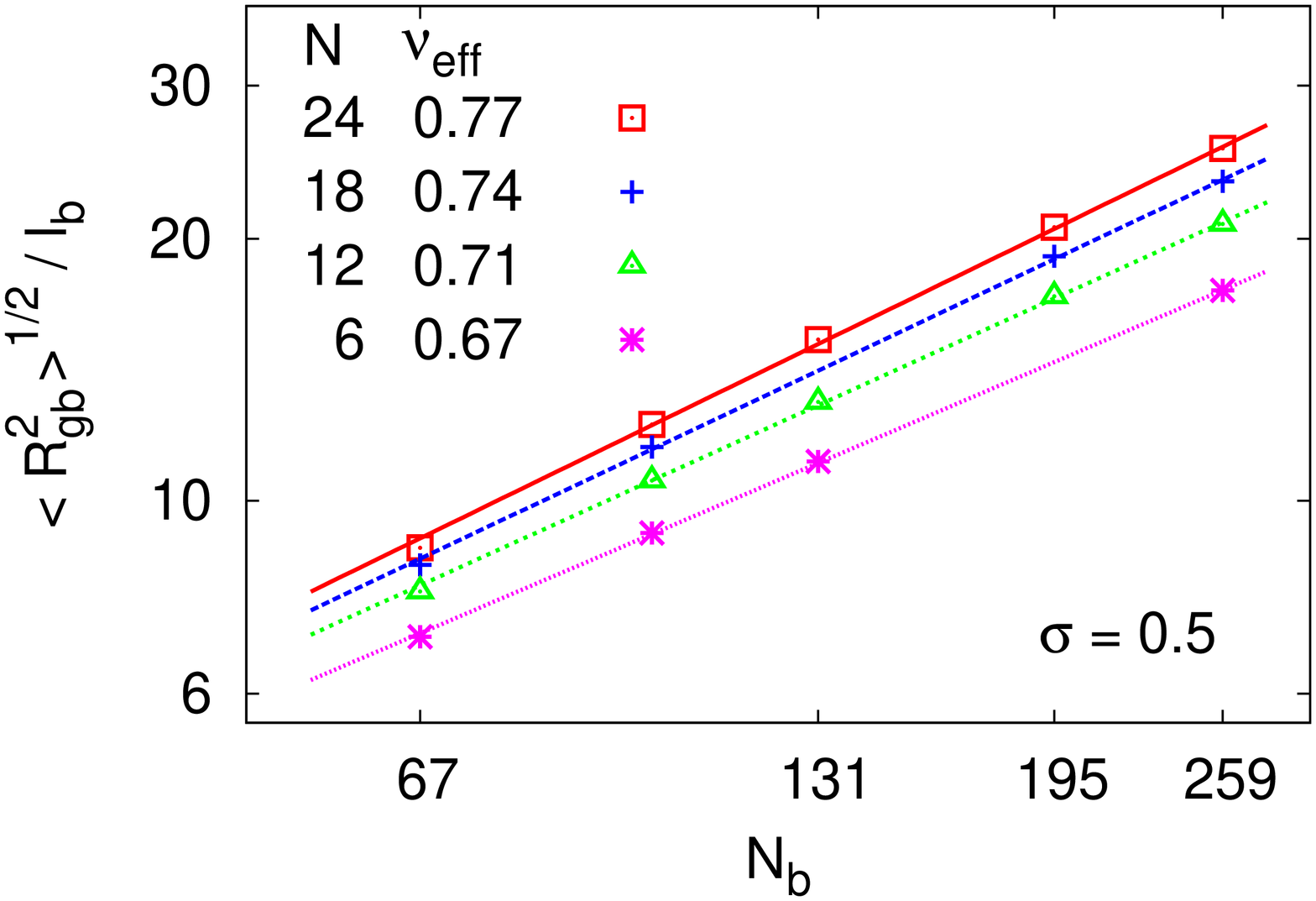}\hspace{0.8truecm}
(b)\includegraphics[scale=0.28,angle=270]{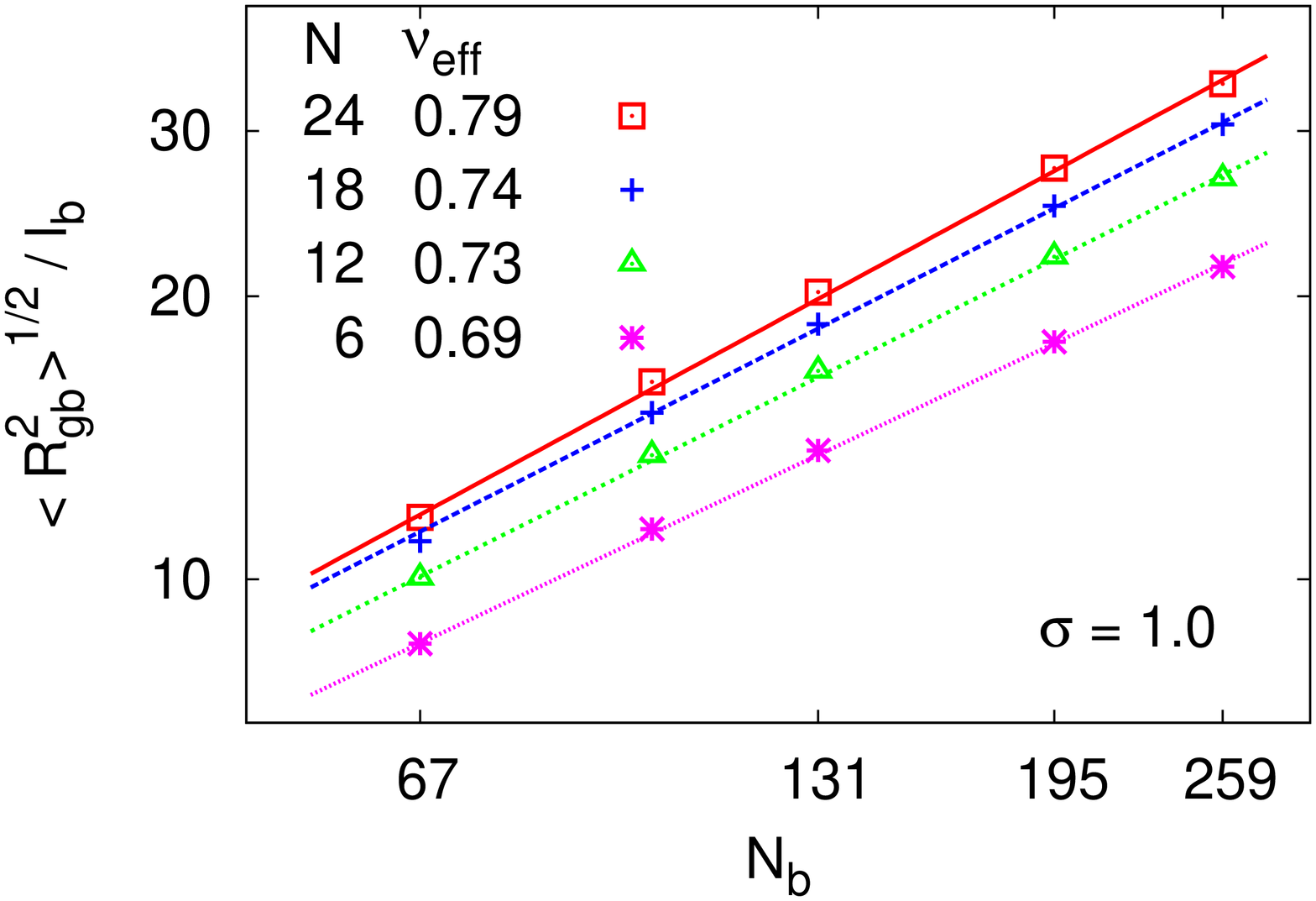}\\
(c)\includegraphics[scale=0.28,angle=270]{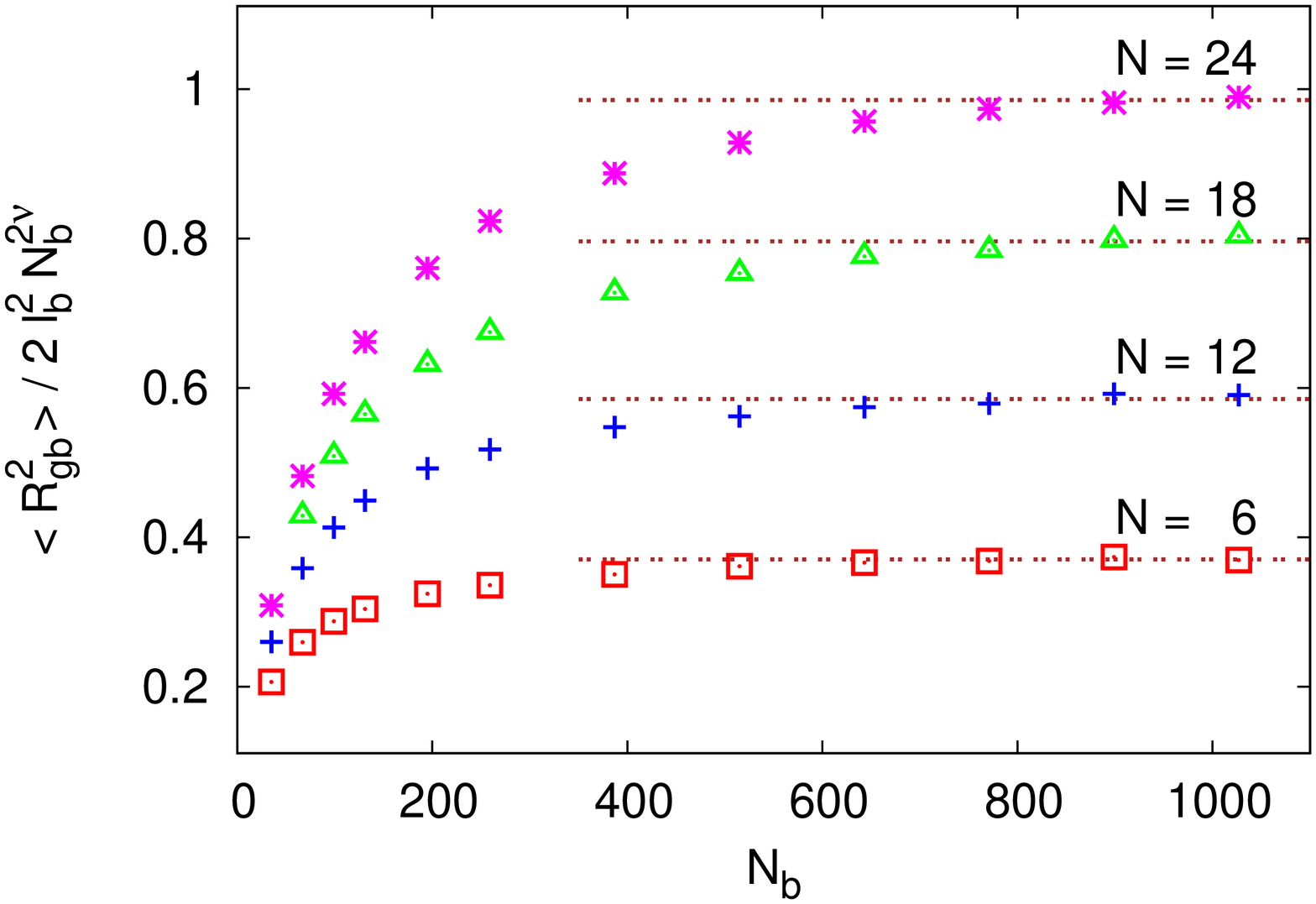}\hspace{0.8truecm}
(d)\includegraphics[scale=0.28,angle=270]{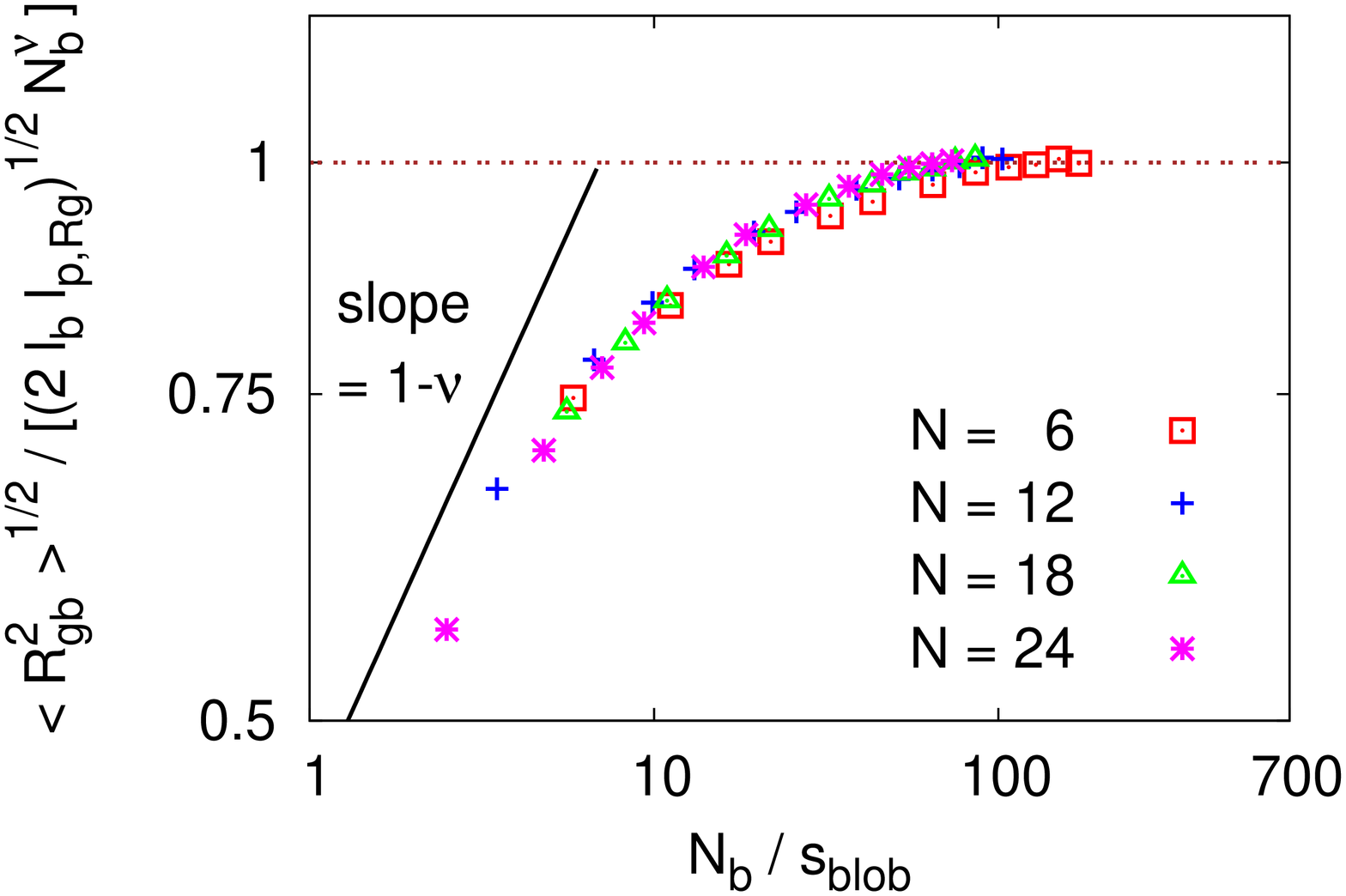}\\
\caption{Log-log plot of 
{$\langle R_{gb}^2 \rangle^{1/2}/\ell_b$} 
for the bond fluctuation model versus
backbone chain length $N_b$, for $67 \le N_b \le 259$,
and several side chain lengths $N$, for $\sigma=0.5$ (a)
and $\sigma=1.0$ (b). Part (c) shows the plot of rescaled
radius of gyration $\langle R^2_{gb} \rangle / (2l_b^2 N_b^{2\nu})$
of the bottle brush polymers versus $N_b$ for $\sigma=1.0$.
The persistence length $l_{p,Rg}$ is determined by the values of plateau.
Part (d) shows a crossover scaling plot,
collapsing for $N=6$, $12$, $18$, and $24$, $\sigma=1.0$, and all data for 
$67\le N_b\le1027$
on a master curve, that describes the crossover from rods
$\left(\langle R_{gb}^2 \rangle^{1/2} \propto N_b \right)$ to swollen
coils $\left( \langle R_{gb}^2 \rangle^{1/2} \propto N_b^{\nu}
\;,\,\nu \approx 0.588 \right)$. For this purpose, $N_b$ is rescaled with
the blob diameter $s_{\rm blob}$, which has been determined to be
$s_{\rm blob}=6$, $10$, $12$, and $14$ for $N=6$, $12$, $18$, and $24$,
respectively~\cite{6,11}.}
\label{fig9}
\end{center}
\end{figure}

   A further effective exponent describes the variation of the 
backbone end-to-end distance and gyration radius with backbone chain length 
(Figs.~\ref{fig8}, \ref{fig9}). We have interpreted this variation as
a crossover from rod-like behavior at small $N_b$ to self-avoiding walk-like
behavior (under good solvent conditions) or random walk-like behavior
( under Theta conditions) with increasing $N_b$.

\begin{figure}
\begin{center}
(a)\includegraphics[scale=0.28,angle=270]{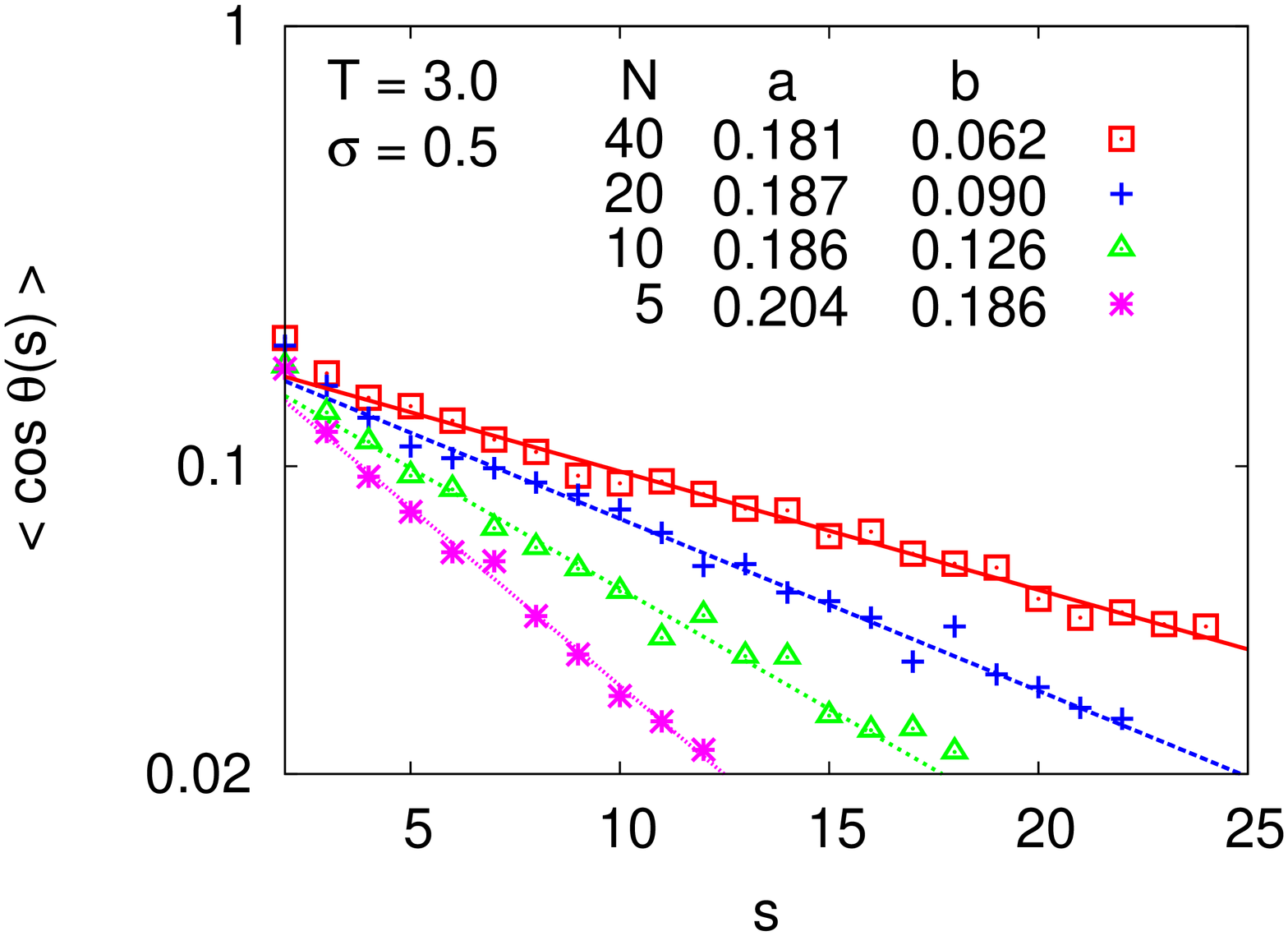}\hspace{0.8truecm}
(b)\includegraphics[scale=0.28,angle=270]{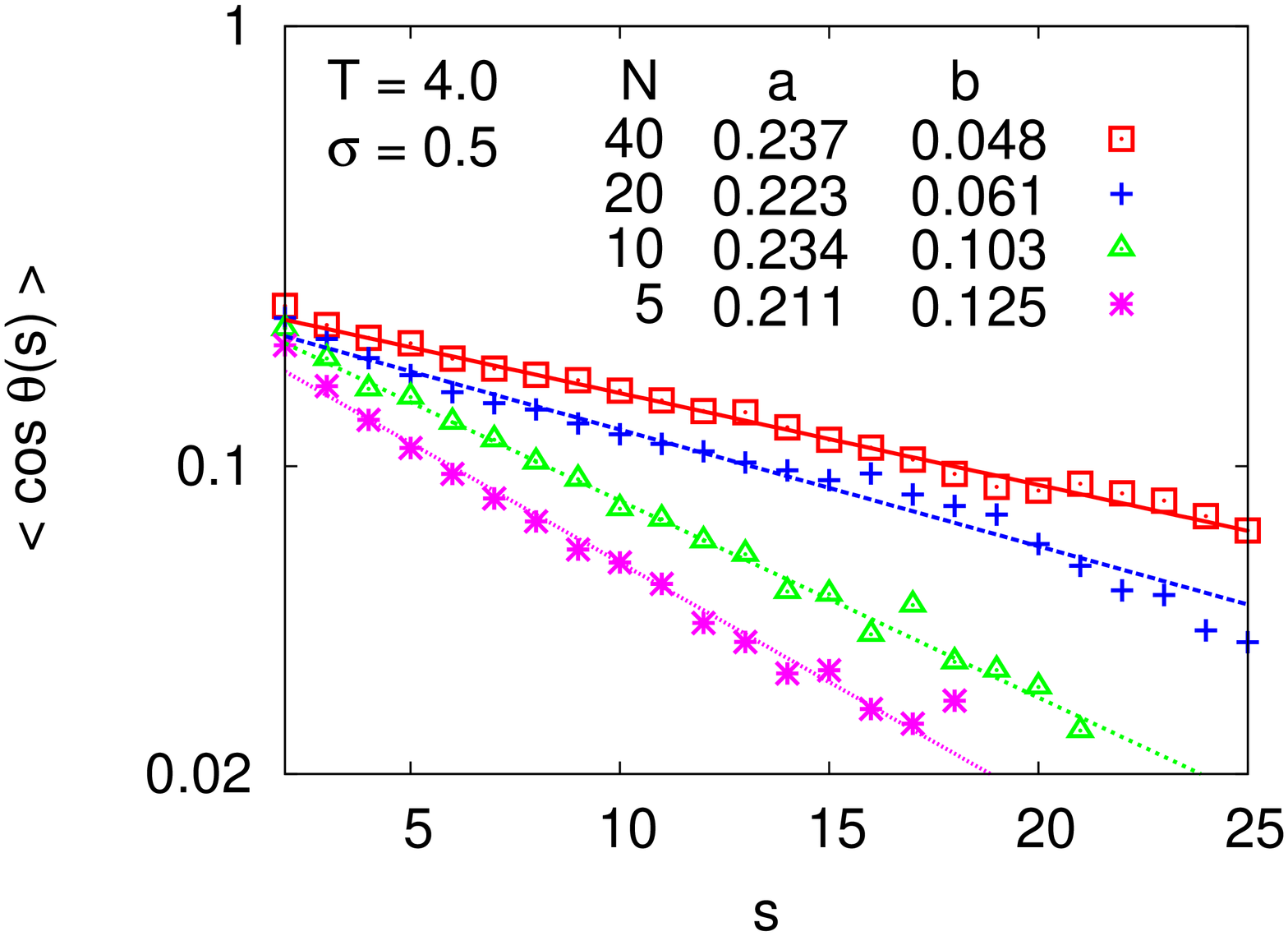}\\
(c)\includegraphics[scale=0.28,angle=270]{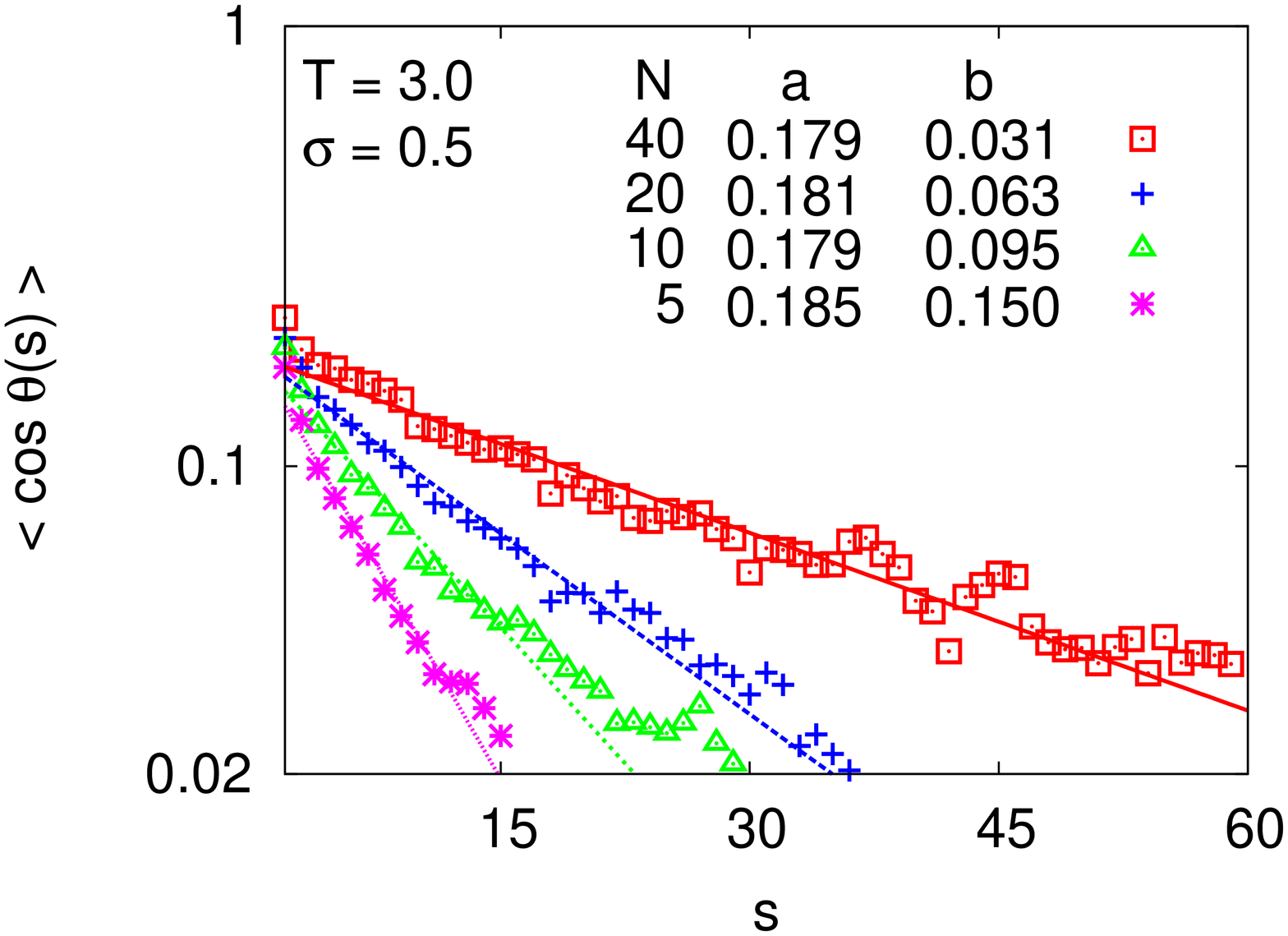}\hspace{0.8truecm}
(d)\includegraphics[scale=0.28,angle=270]{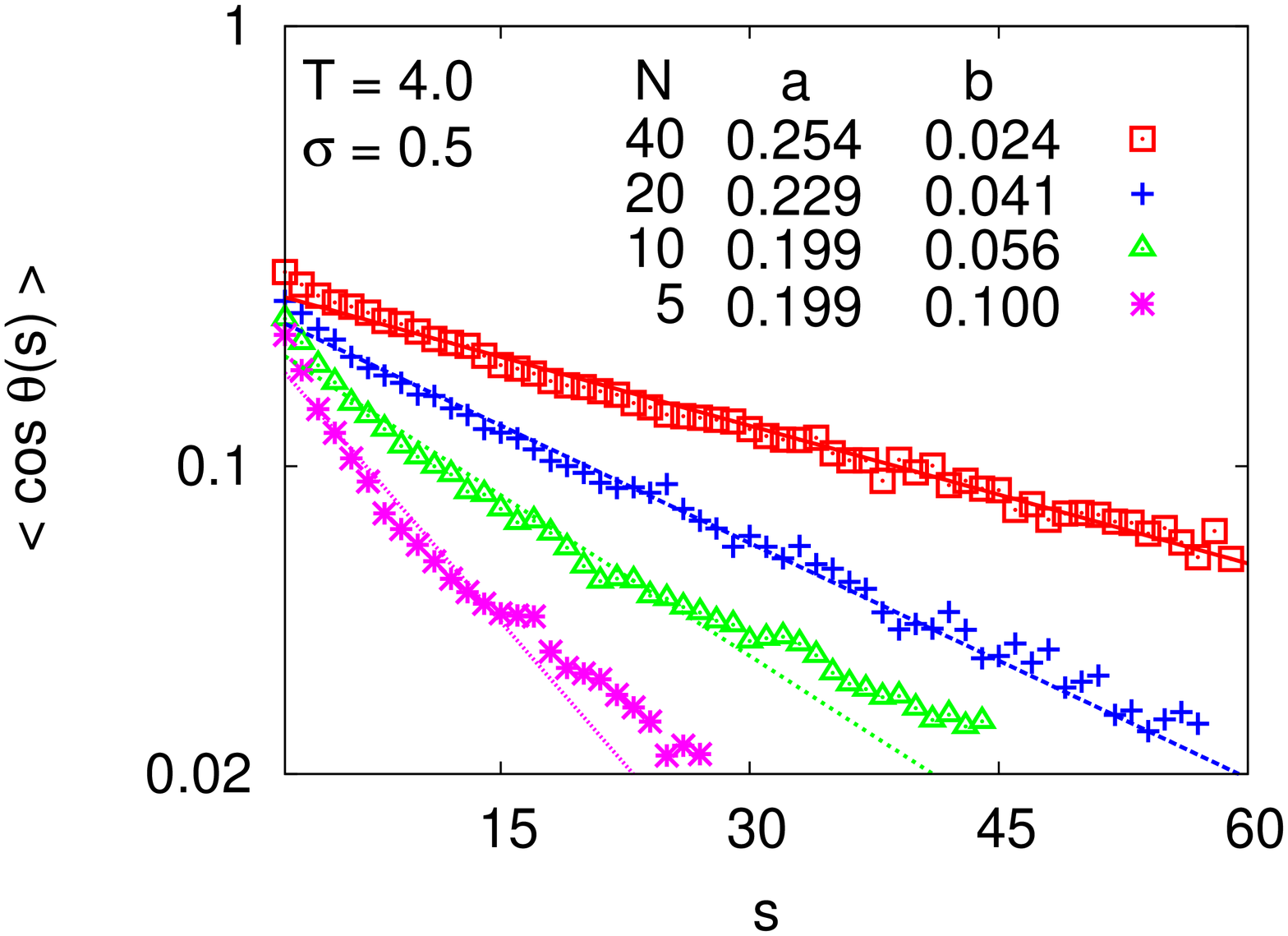}\\
\caption{Semi-log plot of $\langle \cos \theta (s) \rangle$ vs. $s$ for
the bead spring model with $\sigma=0.5$. Data are chosen for  
$N_b=50$ (a)(b) and $N_b=200$ (c)(d) at $T=3.0$
(a)(c) and $T=4.0$ (b)(d), respectively. Several choices of $N$ are shown
as indicated.
Straight lines illustrate fits to  $a\exp(-bs)$, with constants $a$, $b$ 
quoted in the figure.}
\label{fig10}
\end{center}
\end{figure}

\begin{figure}
\begin{center}
(a)\includegraphics[scale=0.28,angle=270]{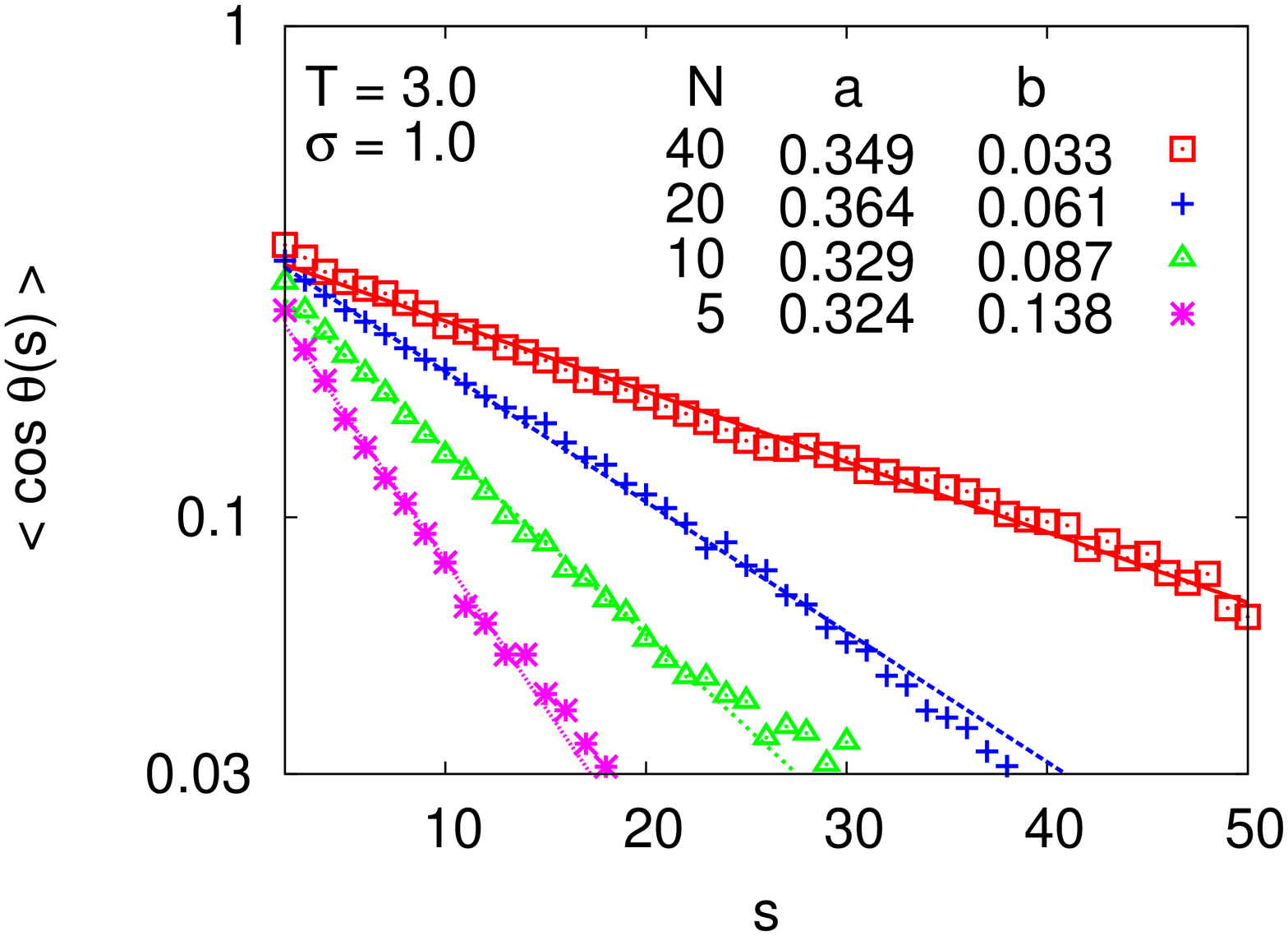}\hspace{0.8truecm}
(b)\includegraphics[scale=0.28,angle=270]{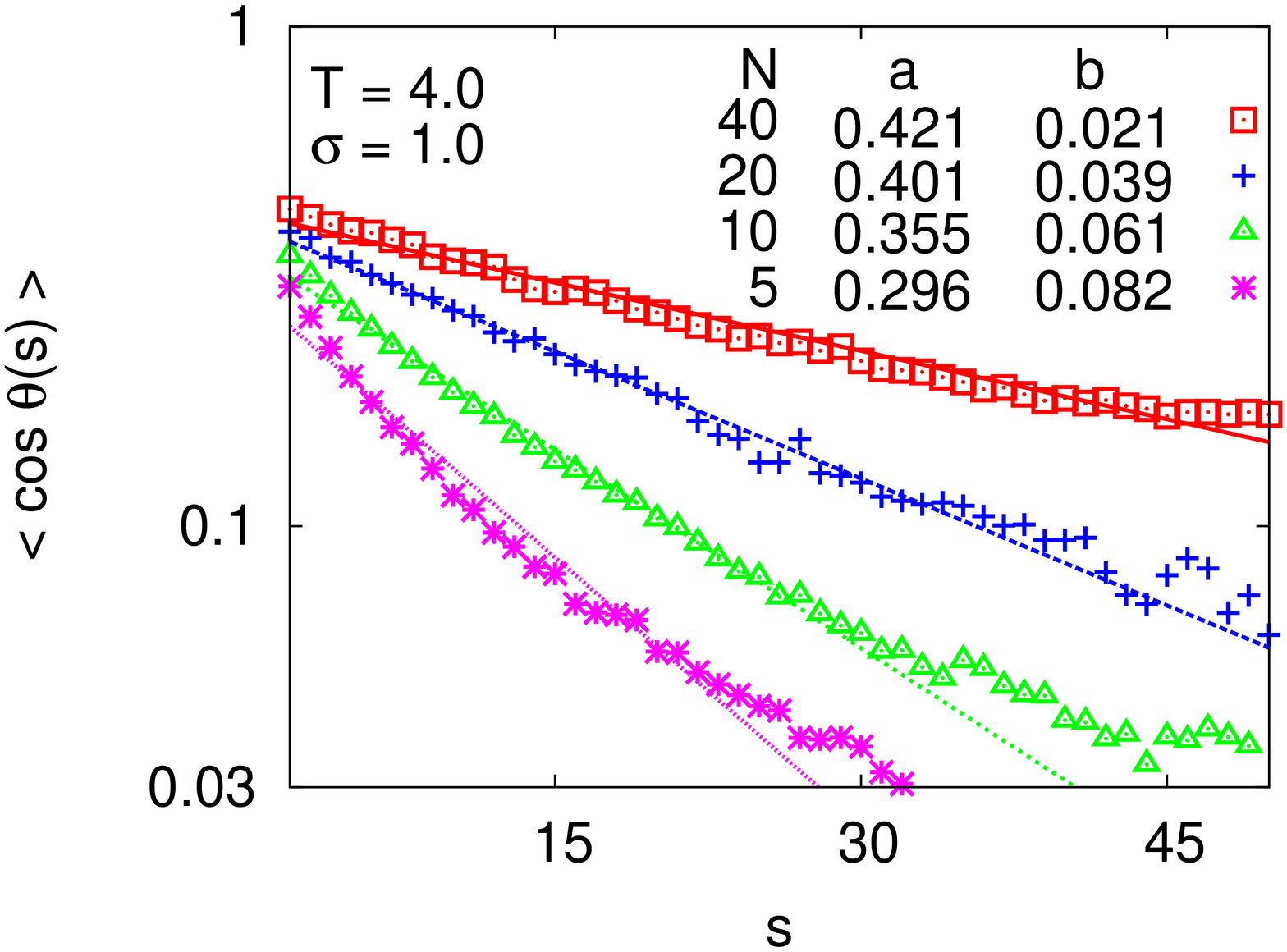}\\
\caption{Semi-log plot of $\langle \cos \theta (s) \rangle$ vs. $s$ for
the bead spring model with $\sigma=1.0$ and $N_b=100$
at the temperatures $T=3.0$ (a) and $T=4.0$ (b),
respectively. Several choices of $N$ are shown as indicated. 
Straight lines illustrate 
fits to a $\exp(-bs)$, with constants $a$, $b$ as quoted in the figure.}
\label{fig11}
\end{center}
\end{figure}

\begin{figure}
\begin{center}
(a)\includegraphics[scale=0.28,angle=270]{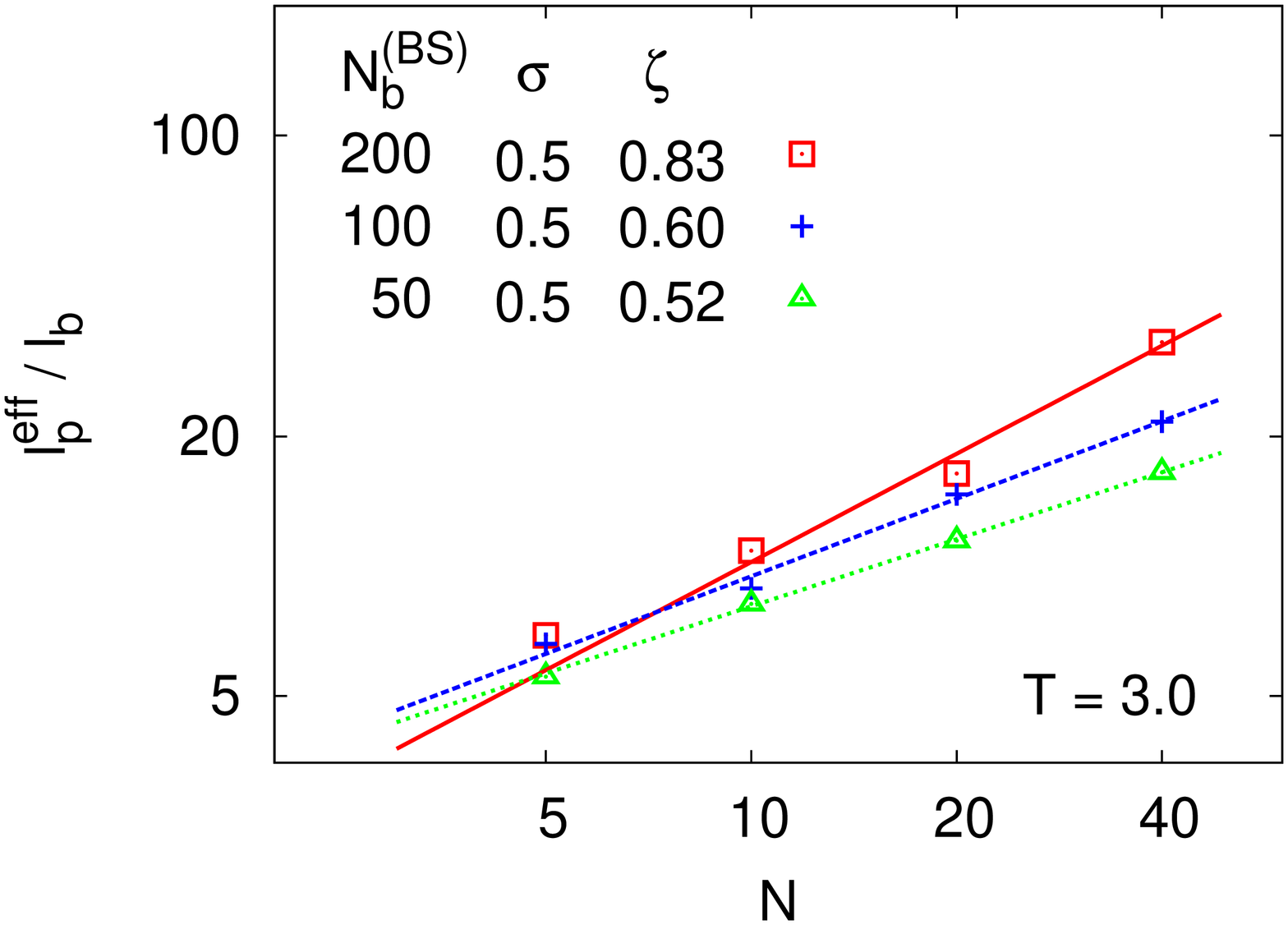}\hspace{0.8truecm}
(b)\includegraphics[scale=0.28,angle=270]{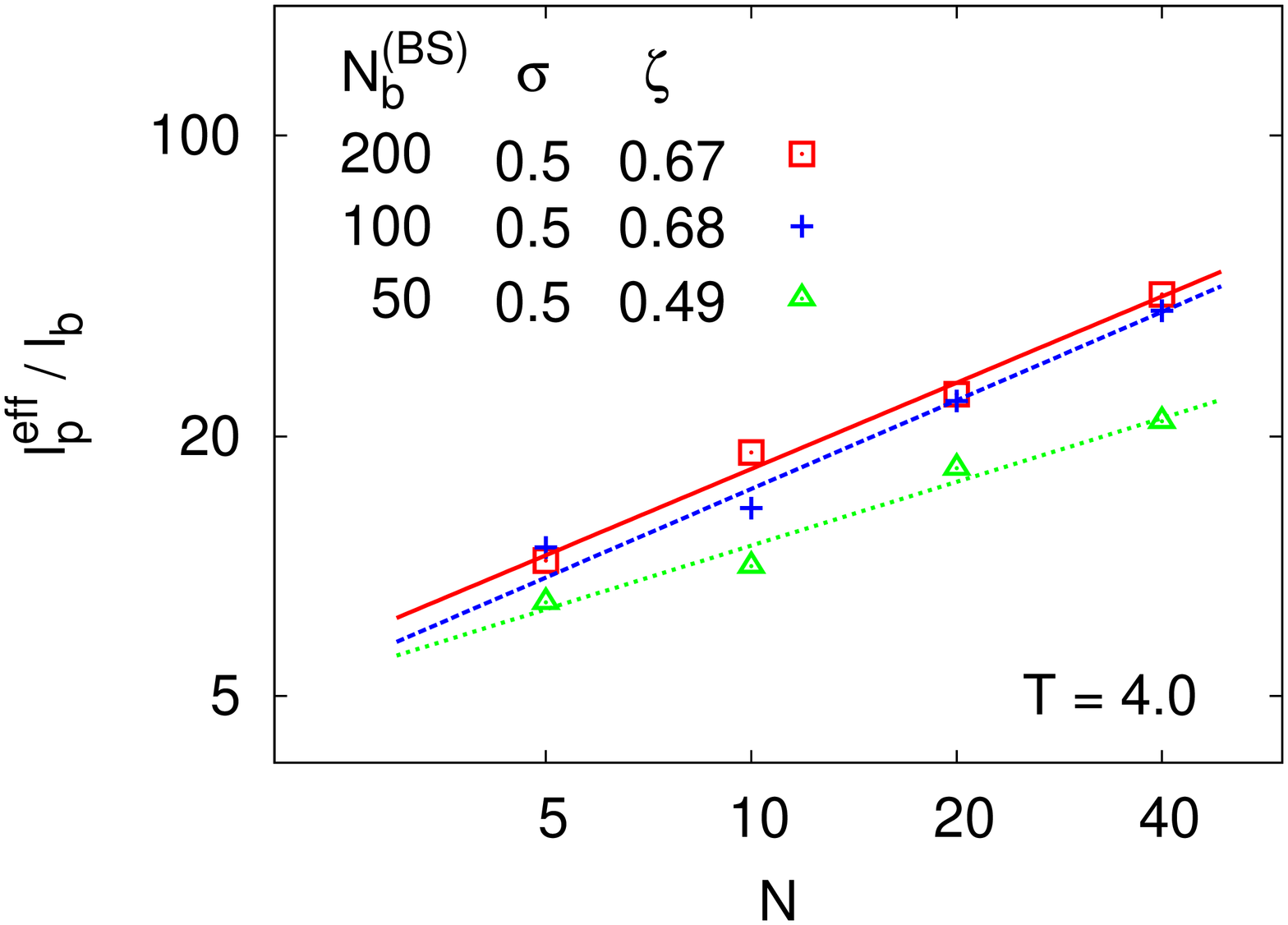}\\
(c)\includegraphics[scale=0.28,angle=270]{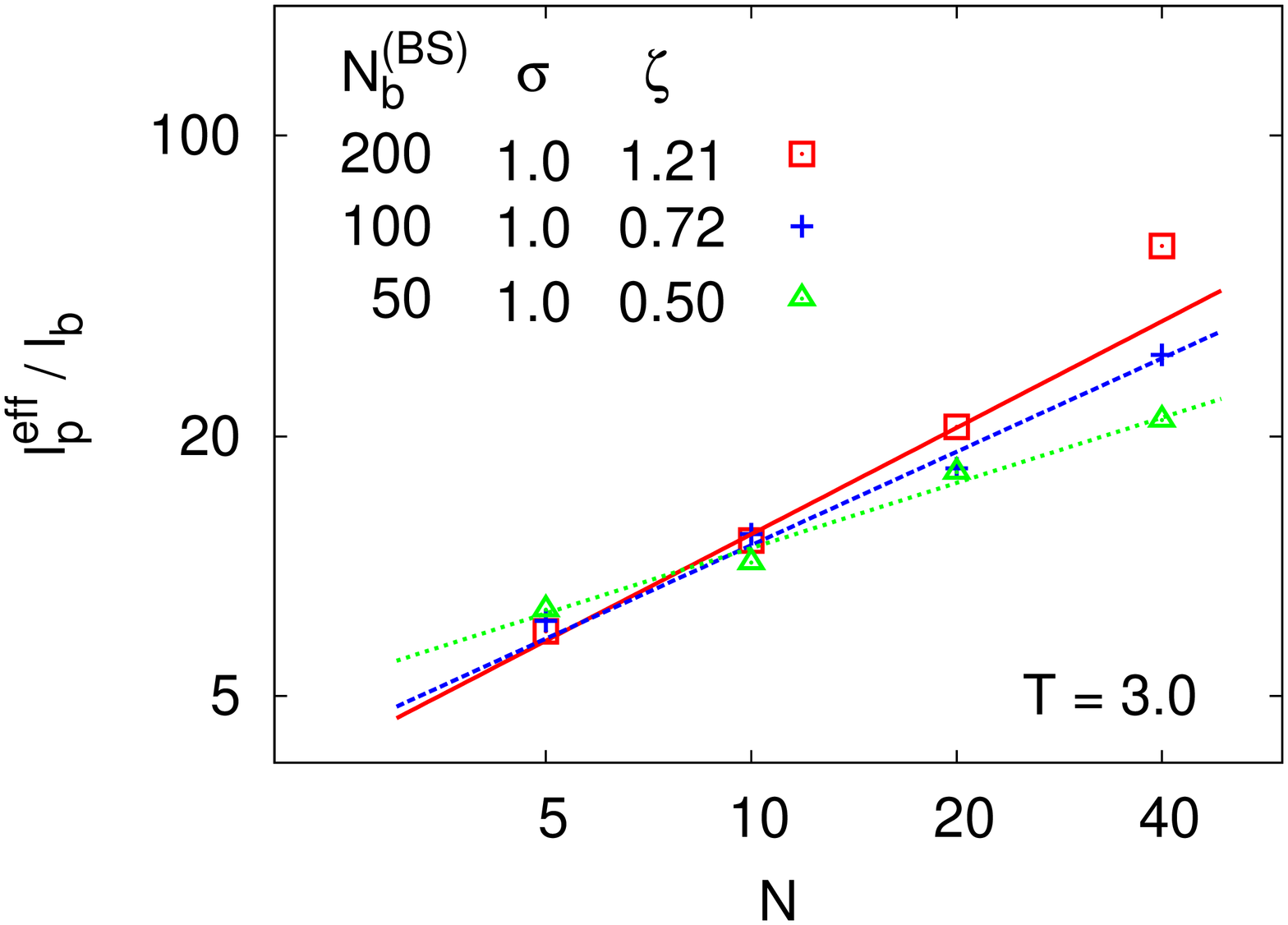}\hspace{0.8truecm}
(d)\includegraphics[scale=0.28,angle=270]{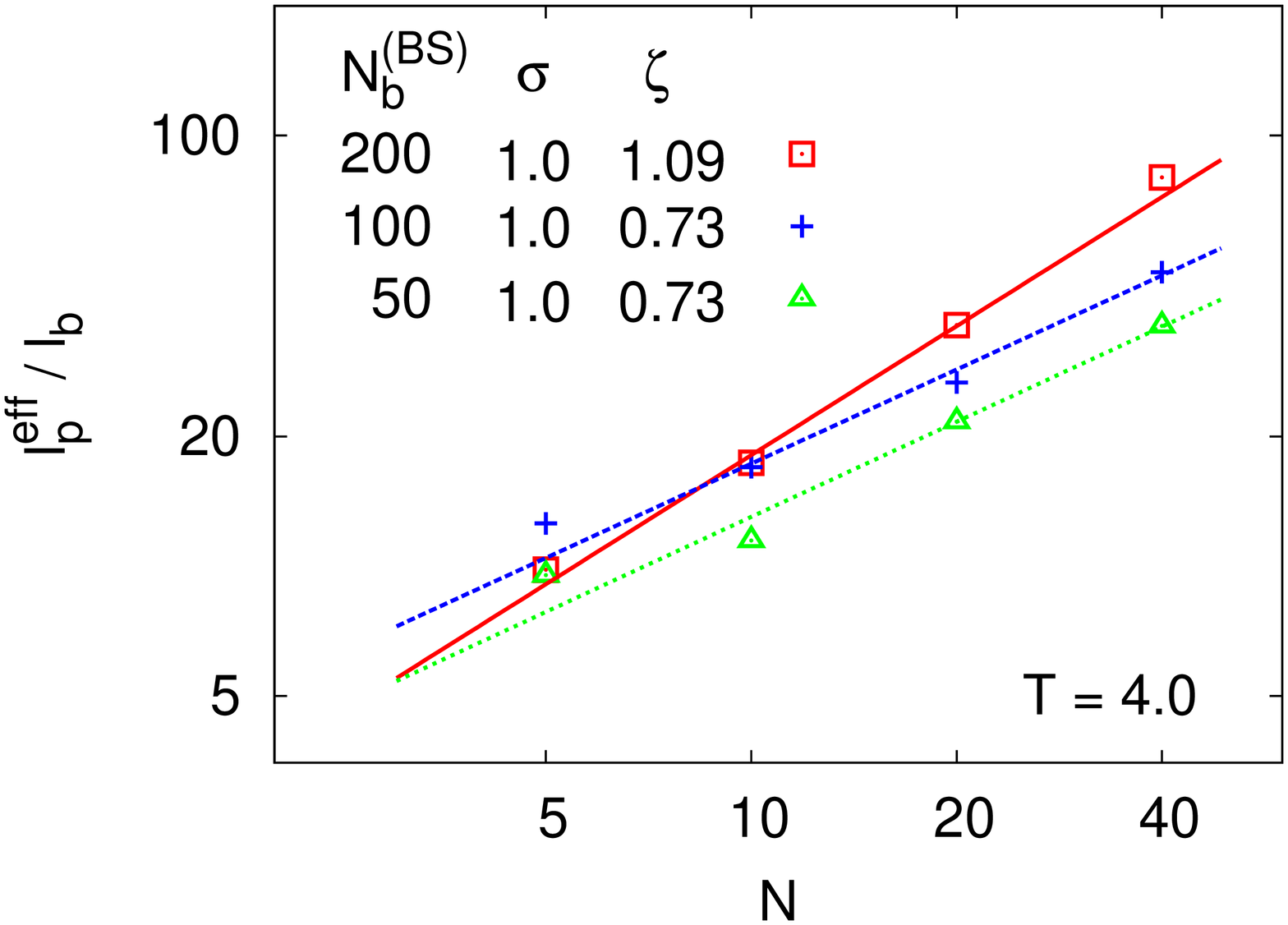}\\
\caption{Log-log plot of the effective persistence length 
{$l_p^{\rm eff}/\ell_b=b^{-1}$}
as a function of side chain length $N$
for $T=3.0$ (a)(c) and $T=4.0$ (b)(d) for several choices of $N_b$.
Results for $\sigma=1.0$ and $\sigma=0.5$ are shown as indicated
{and the exponent estimates $\zeta$ extracted from the slope are listed}.
Values of $l_p^{\rm eff}$ are also listed in Table~\ref{table1}.}
\label{fig12}
\end{center}
\end{figure}

\begin{figure}
\begin{center}
(a)\includegraphics[scale=0.28,angle=270]{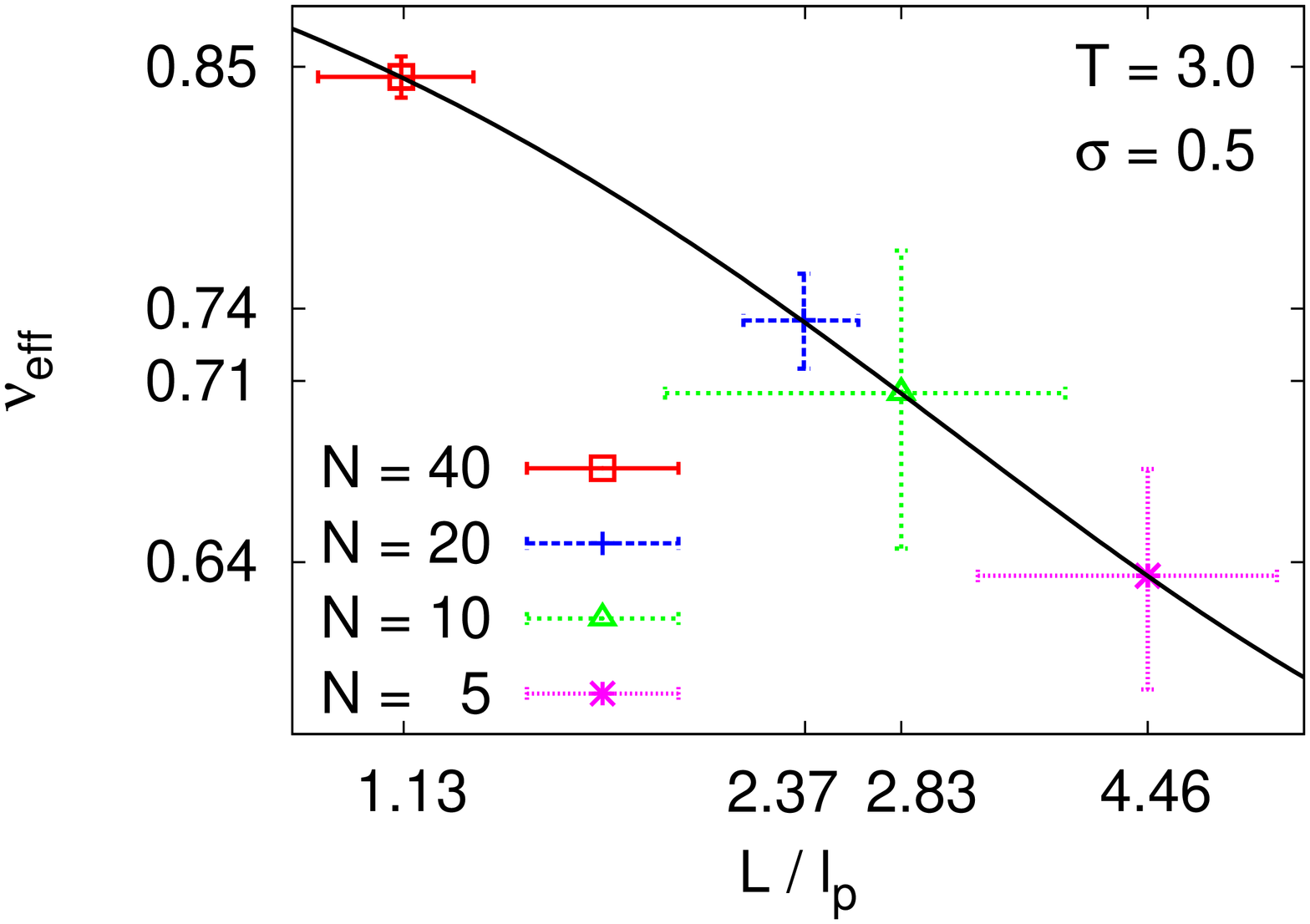}\hspace{0.8truecm}
(b)\includegraphics[scale=0.28,angle=270]{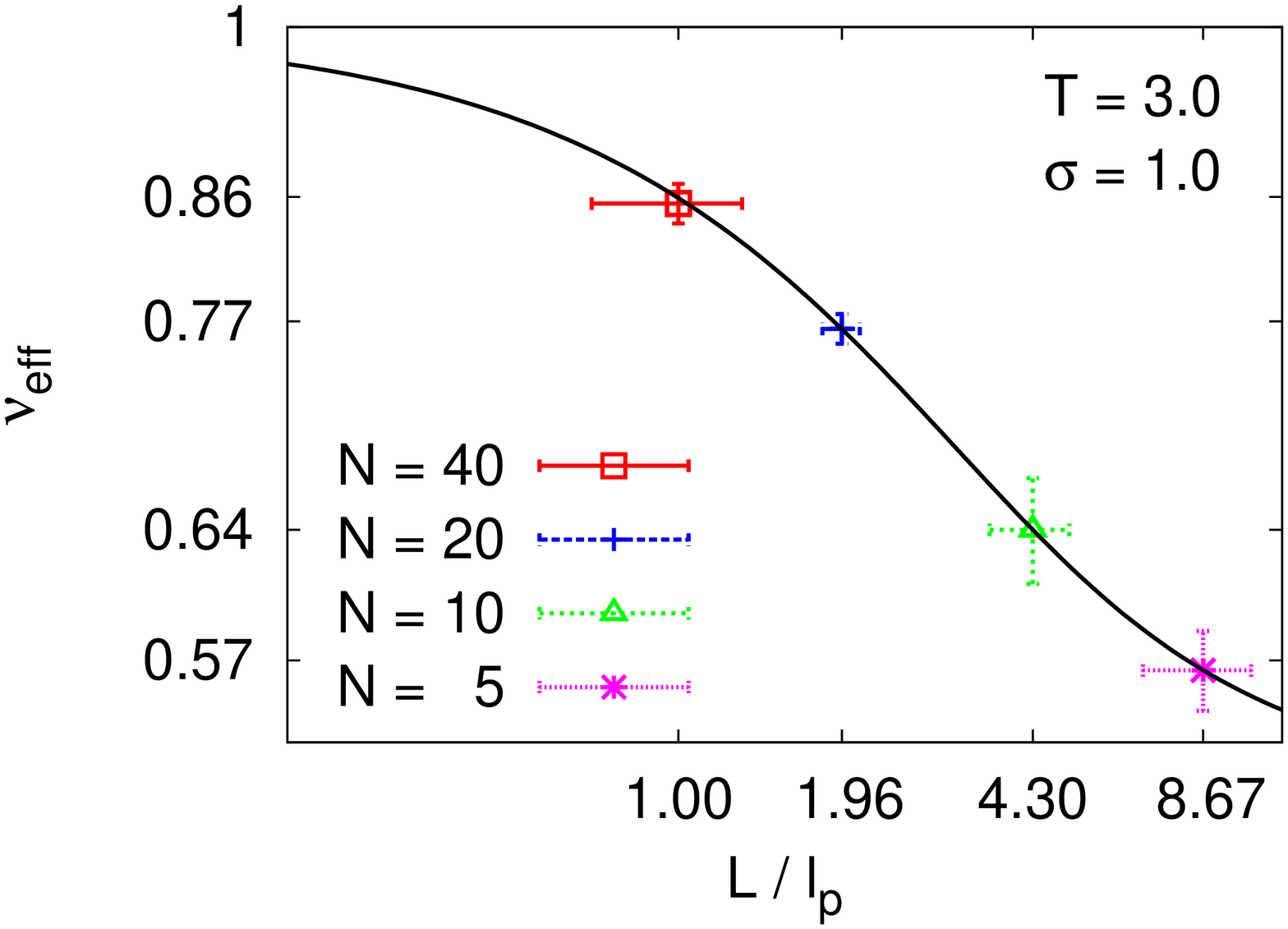}\\
(c)\includegraphics[scale=0.28,angle=270]{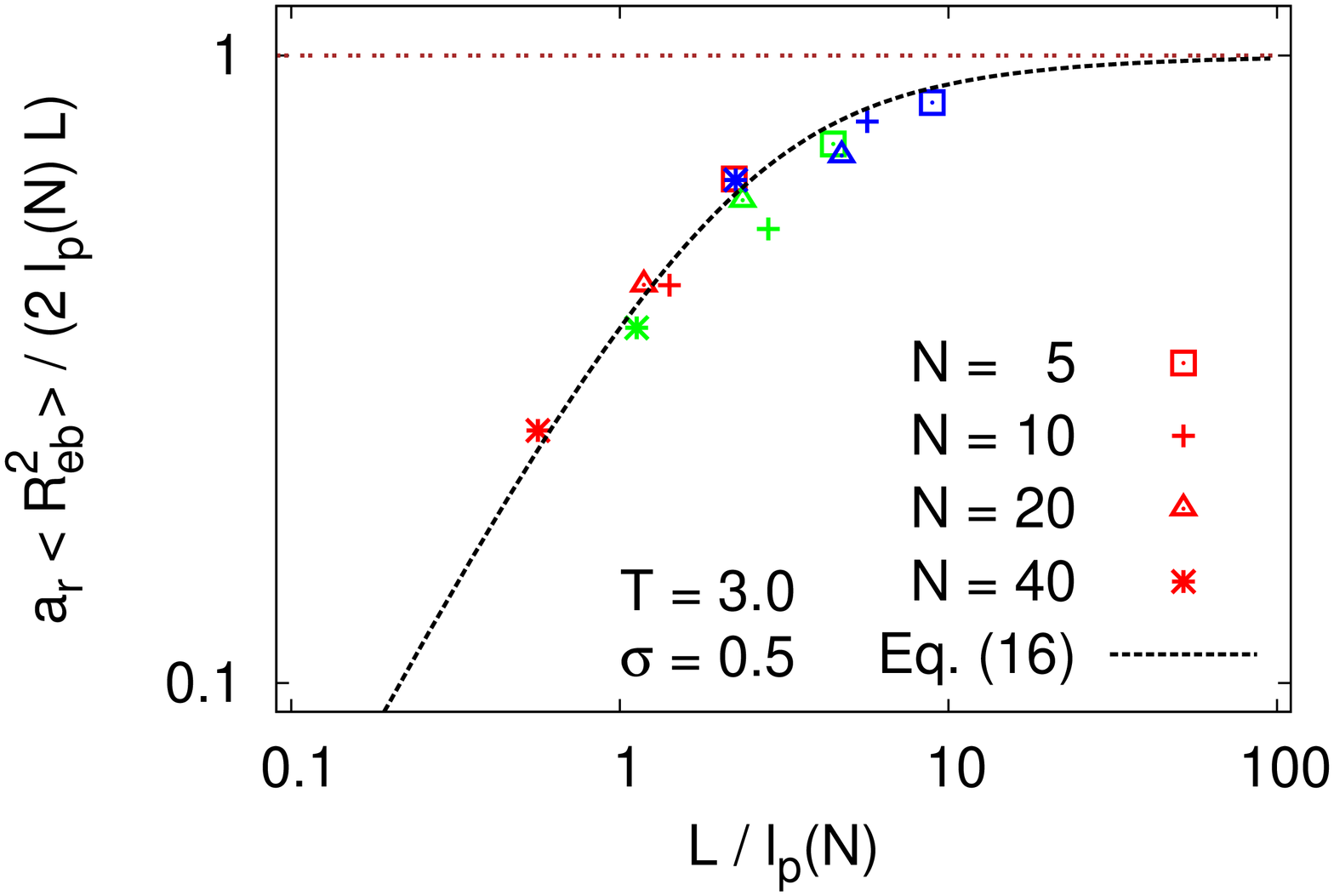}\hspace{0.8truecm}
(d)\includegraphics[scale=0.28,angle=270]{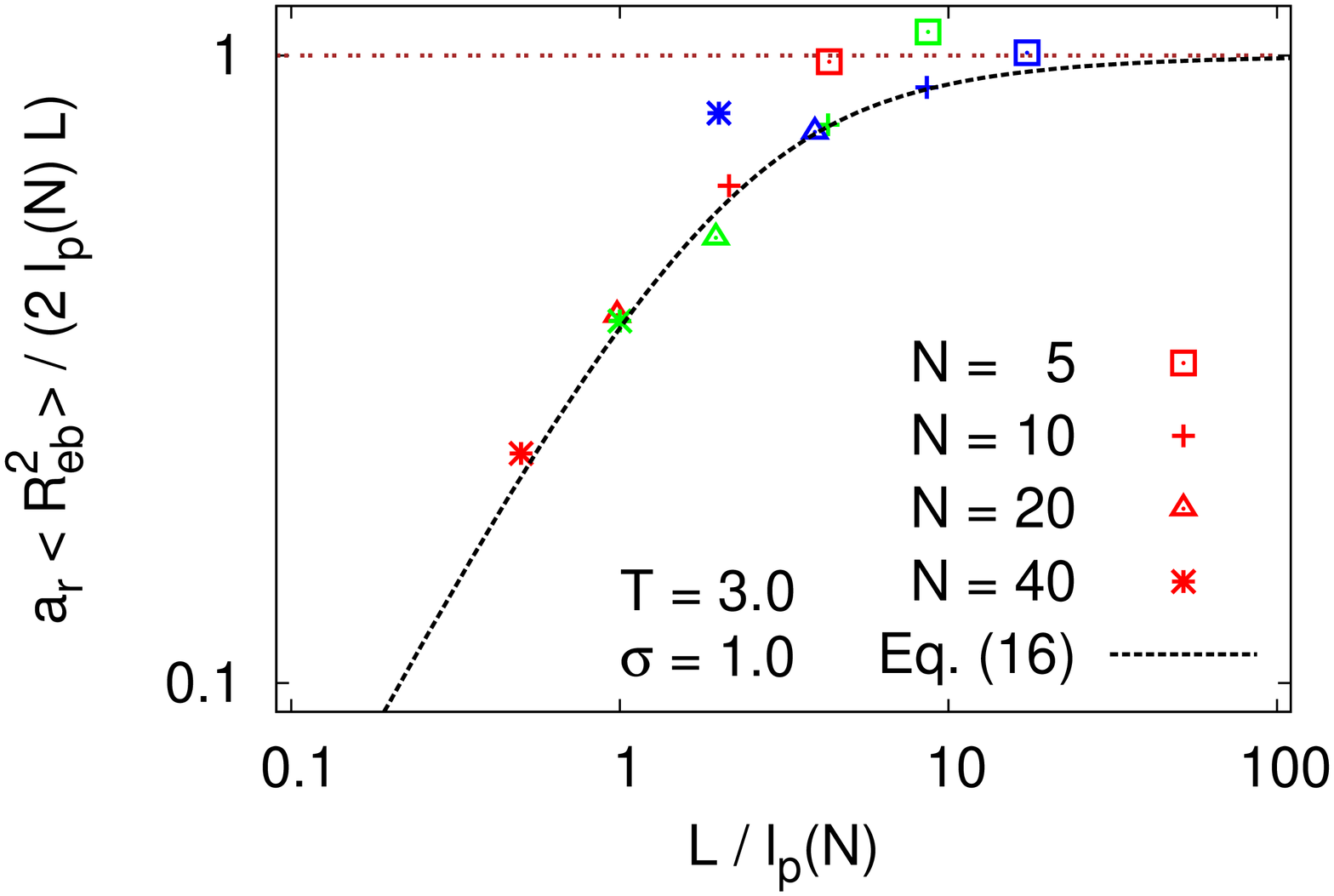}\\
\caption{Plot of the function $\nu_{\rm eff}(L/l_p)$ versus $L/l_p$
as predicted from the Kratky-Porod model, Eqs.~(\ref{eq16}), (\ref{eq17}) (full
curves).
The numbers for $L/l_p$ extracted for $\sigma=0.5$ (a) and $\sigma=1.0$ (b)
are quoted in the figure for various $N$
at the abscissa. The average of the exponents $\nu_{\rm eff}$
from Fig.~\ref{fig8}(a)(c) and from $R_{eb}^2$ (not shown) are quoted on the
ordinate. Log-log plot of $a_r\langle R_{eb}^2 \rangle/(2l_p(N)L)$
vs. $L/l_p(N)$ for $\sigma=0.5$ (c) and $\sigma=1.0$ (d). 
Several choices of $N$ are shown as indicated. Data plotted by the
same symbol correspond to $N_b=50$, $100$, and $200$ from left to right.
The prediction for the Kratky-Porod model \{Eq.~(\ref{eq16})\} is also shown
for comparison. Approximate data collapse is obtained by introducing a factor
$a_r$ ($a_r=9.2\pm1.4.$ and $4.9\pm0.2$ for $\sigma=0.5$ and $1.0$, 
respectively), cf. text. Values of $l_p(N)$ are also listed in 
Table~\ref{table1}.} 
\label{fig13}
\end{center}
\end{figure}

   Finally, ``the'' persistence length has been estimated in various ways,
and demonstrating that the different estimates are not mutually consistent
with each other it was argued that a unique persistence length does not
exist. While a regime of exponential decay of the bond autocorrelation function
often can be observed over some intermediate range of contour distances
(Fig.~\ref{fig10}, \ref{fig11}), the resulting estimates for a persistence
length do not only depend on side chain length $N$ and grafting density 
$\sigma$, but also on backbone chain length $N_b$ and on temperature $T$
(Fig.~\ref{fig12}). However, when one studies the variation of the
end-to-end distance of the backbone for $T=T_\theta$ an analysis in 
terms of the 
Kratky-Porod wormlike chain model \{Eq.~(\ref{eq16})\} becomes 
feasible (Fig.~\ref{fig13}). But one must not identify the contour length
$L$ implied by this model with the ``chemical'' contour length
$L_{\rm ch} = N_b l_b$, where $l_b$ is the actual bond length,
but rather one has $L=N_b l_b^{\rm eff}$ with $l_b^{\rm eff}$
distinctly smaller than $l_b$. This effect results from the 
flexibility of the backbone on small scales; only on the scale of
several backbone bonds does the stiffening due to the mutual side chain 
repulsions come into play.
{
Thus, at the Theta point both an effective contour length $L$ and
a persistence length $\ell_p(N)$ are well-defined quantities, in terms
of a fit of the data to the Kratky-Porod model, while under good solvent
conditions such an analysis is not appropriate.}

  We hope that the present work stimulates some more systematic experimental
work on these issues. Of course, for real bottle brushes where the chemical
nature of the backbone often differs from the side chains, the interesting
possibility arises that the backbone and the side chains have rather different 
Theta temperatures. Another subtle issue concerns the precise location
of Theta conditions. While we expect that for the limit 
$N_b \rightarrow \infty$ at fixed ratio $N/N_b$ the Theta conditions 
coincide with the Theta conditions for simple linear polymers
($N_b \rightarrow \infty$, $N=0$), it is not clear whether the location of 
the Theta point for the limit $N_b \rightarrow \infty$ at fixed
(small) $N$ depends significantly on $N$. Thus, also more theoretical
work clearly is needed.

\begin{acknowledgments} P.E.T. thanks for financial support by
the Austrian Science Foundation within the SFB ViCoM (grant F41), and
also the Max Planck Fellowship during the time he was in Mainz.
H.-P. H. received funding from the
Deutsche Forschungsgemeinschaft (DFG), grant No SFB 625/A3. We are
grateful for extensive grants of computer time at the JUROPA 
under the project No HMZ03 and
SOFTCOMP computers at the J\"ulich Supercomputing Centre (JSC).
\end{acknowledgments}

\clearpage
\begin{table}[htb]{\caption{Estimates of the effective persistence length 
{$l_p^{\rm eff}=b^{-1}\ell_b$} shown in Fig.~\ref{fig12}
are listed for the BS model with the grafting densities $\sigma=0.5$ and $1.0$ 
at temperatures $T=3.0$ and $4.0$. Various values of the backbone length $N_b$
and the side chain length $N$ are chosen here. 
{All length quoted in this table are given in units of $\ell_b$
(with $\ell_b \approx 0.97 \sigma_{\rm LJ}$).
Note that $\ell_p^{\rm eff}$ depends not only on $N$ and $\sigma$,
but also on $N_b$ and $T$, and hence does not seem as a characteristic
of intrinsic chain stiffness.}
Estimates of $l_p(N)$ shown
in Fig.~\ref{fig13} are also listed for comparison.}
\label{table1}}
\begin{tabular}{|rr|rr|rr|}
\hline
\multicolumn{2}{|l|}{} &\multicolumn{2}{c|}{$\sigma=0.5$} & \multicolumn{2}{c|}{$\sigma=1.0$} \\
\hline
$N_b$ & $N$ & ${l_p^{\rm eff}/\ell_b}(T=3.0)$ & 
${l_p^{\rm eff}/\ell_b}(T=4.0)$ & 
${l_p^{\rm eff}/\ell_b}(T=3.0)$ & 
${l_p^{\rm eff}/\ell_b}(T=4.0)$ \\ 
\hline
50 &  5  &  5.38  &  8.00  &  7.69  &  9.26 \\
50 & 10  &  7.94  &  9.71  &  9.90  & 11.14 \\
50 & 20  & 11.12  & 16.39  & 16.05  & 21.01 \\
50 & 40  & 16.03  & 21.27  & 21.19  & 35.09 \\
\hline
100 &  5 &   6.41 &  10.73 &   7.25 &  12.20 \\
100 & 10 &   8.26 &  13.33 &  11.52 &  16.47 \\
100 & 20 &  14.25 &  23.47 &  16.37 &  25.91 \\
100 & 40 &  21.01 &  38.46 &  38.03 &  47.62 \\
\hline
200 &  5 &   6.71 &  10.00 &   6.85  &   9.52 \\
200 & 10 &  10.55 &  17.83 &  11.14  & 16.95 \\
200 & 20 &  15.92 &  24.33 &  20.45  & 35.21 \\
200 & 40 &  32.15 &  41.49 &  53.76  & 77.52 \\
\hline
\hline
& N & ${l_p/\ell_b}(T=3.0)$ &  & 
${l_p/\ell_b}(T=3.0)$ &   \\
\hline
& 5 & 23.11  &    &  11.89 &  \\
&10 & 36.43  &    &  23.98 &  \\
&20 & 43.59  &    &  52.60 &  \\
&40 & 91.64  &    & 103.09 &  \\
\hline
\end{tabular}
\end{table}

\clearpage

\end{document}